\documentclass[superscriptaddress,twocolumn,prb,aps,showpacs,floatfix]{revtex4}

\usepackage{epsfig}
\usepackage{epsf}
\usepackage{graphicx}
\usepackage{amsmath}
\usepackage{booktabs}  
\usepackage{dcolumn}
\usepackage{rotating}
\usepackage{bm}        
\usepackage{bbm}       
\newcommand{\be}{\begin{equation}}
\newcommand{\ee}{\end{equation}}
\newcommand{\bea}{\begin{eqnarray}}
\newcommand{\eea}{\end{eqnarray}}
\newcommand{\ba}{\begin{array}}
\newcommand{\ea}{\end{array}}
\newcommand{\mint}[1]{\int\! \D^{3} #1 \, }
\newcommand{\mdint}[2]{\mint{#1}\!\!\!\mint{#2}}

\newcommand{\I}{{\rm i}}
\newcommand{\D}{{\rm d}}
\newcommand{\E}{{\rm e}}

\newcommand{\bk}{{\bm{k}}}
\newcommand{\bem}{{\bm{m}}}
\newcommand{\bp}{{\bm{p}}}
\newcommand{\bq}{\bm{q}}
\newcommand{\bQ}{\bm{Q}}
\newcommand{\br}{\bm{r}}
\newcommand{\bR}{\bm{R}}
\newcommand{\bs}{\bm{s}}

\newcommand{\bx}{\bm{x}}

\newcommand{\bek}[1]{\left(\frac{\beta}{2}{#1}\right)}

\newcommand{\up}{\uparrow}
\newcommand{\down}{\downarrow}

\newcommand{\op}[1]{{\hat #1}}
\newcommand{\Tr}[1]{{\rm Tr}\{ #1 \}}
\newcommand{\KS}{{\rm s}}

\begin{document}

\title{Ab-initio theory of superconductivity -- I: Density functional formalism and
approximate functionals}

\author{M.~L{\"u}ders}
\affiliation{Daresbury Laboratory, Warrington WA4 4AD, United Kingdom }
\affiliation{Institut~f\"ur Theoretische Physik, Universit\"at 
W\"urzburg, Am Hubland, D-97074 W\"urzburg, Germany}

\author{M.\,A.\,L.~Marques}
\author{N.\,N.~Lathiotakis}
\affiliation{Institut f{\"u}r Theoretische Physik, Freie Universit{\"a}t Berlin, Arnimallee 14, D-14195 Berlin, Germany} 
\affiliation{Institut~f\"ur Theoretische Physik, Universit\"at 
W\"urzburg, Am Hubland, D-97074 W\"urzburg, Germany}

\author{A.~Floris}
\affiliation{Institut f{\"u}r Theoretische Physik, Freie Universit{\"a}t Berlin, Arnimallee 14, D-14195 Berlin, Germany} 
\affiliation{INFM SLACS, Sardinian Laboratory for Computational Materials Science and
Dipartimento di Scienze Fisiche, Universit\`a degli Studi di Cagliari,
S.P. Monserrato-Sestu km 0.700, I--09124 Monserrato (Cagliari), Italy}

\author{G.~Profeta}
\affiliation{CASTI - Istituto Nazionale Fisica della Materia
(INFM) and Dipartimento di Fisica, Universit\`a degli studi dell'Aquila,
I-67010 Coppito (L'Aquila) Italy}

\author{L.~Fast}
\affiliation{SP Swedish National Testing and Research
Institute, P.O.B. 857, S-501 15 Bor{\aa}s, Sweden}
\affiliation{Institut~f\"ur Theoretische Physik, Universit\"at 
W\"urzburg, Am Hubland, D-97074 W\"urzburg, Germany}

\author{A.~Continenza}
\affiliation{CASTI - Istituto Nazionale Fisica della Materia
(INFM) and Dipartimento di Fisica, Universit\`a degli studi dell'Aquila,
I-67010 Coppito (L'Aquila) Italy}

\author{S.~Massidda}
\altaffiliation{Also at LAMIA-INFM, Genova, Italy}
\affiliation{INFM SLACS, Sardinian Laboratory for Computational Materials Science and
Dipartimento di Scienze Fisiche, Universit\`a degli Studi di Cagliari,
S.P. Monserrato-Sestu km 0.700, I--09124 Monserrato (Cagliari), Italy}
 
\author{E.\,K.\,U.~Gross}
\affiliation{Institut f{\"u}r Theoretische Physik, Freie Universit{\"a}t Berlin, Arnimallee 14, D-14195 Berlin, Germany} 
\affiliation{Institut~f\"ur Theoretische Physik, Universit\"at 
W\"urzburg, Am Hubland, D-97074 W\"urzburg, Germany}

\date{\today}

\begin{abstract}
A novel approach to the description of superconductors in thermal equilibrium is developed within a formally exact
density-functional framework. The theory is formulated in terms of three ``densities'': the ordinary 
electron density, the superconducting order parameter, and the diagonal of the nuclear $N$-body density matrix.
The electron density and the order parameter are determined by Kohn-Sham equations that resemble the
Bogoliubov-de~Gennes equations. The nuclear density matrix follows from a Schr\"odinger equation with 
an effective $N$-body interaction. These equations are coupled to each other via exchange-correlation
potentials which are universal functionals of the three densities. Approximations of these exchange-correlation 
functionals are derived using the diagrammatic techniques of many-body perturbation theory. The bare Coulomb 
repulsion between the electrons and the electron-phonon interaction enter this perturbative treatment on the same footing.
In this way, a truly ab-initio description is achieved which does not contain any empirical parameters.
\end{abstract}


\maketitle

\section{Introduction}
\label{sec:intro}

One of the great challenges of modern condensed-matter theory is the prediction of material 
specific properties of superconductors, such as the critical temperature T$_\text{c}$ or the 
gap at zero temperature $\Delta_0$. The model of Bardeen, Cooper and Scrieffer (BCS)~\cite{bcs} 
successfully describes the universal features of superconductors, i.e. those features which all 
(conventional, weak-coupling) superconductors have in common, like the universal value of the
ratio $2\Delta_0/k_\text{B} T_\text{c}$. The great achievement of BCS theory was the microscopic identification of 
the superconducting order parameter which lead, after more than 50 years of struggling, to a 
microscopic understanding of the phenomenon of superconductivity.

BCS theory, however, cannot be considered a predictive theory in the sense that it would allow
the computation of material-specific properties. Moreover, materials with strong electron-phonon
coupling, such as niobium or lead, are poorly described by BCS theory. In these strong-coupling
materials, phonon retardation effects play a very important role. A proper treatment of those effects
was developed by Eliashberg~\cite{Eliashberg,scalapino}. His theory can be viewed as a GW approximation\cite{GW} in
terms of the Nambu-Gorkov~\cite{pNambu1960} Green's functions. Eliashberg's theory not only 
achieves a successful description of the strong coupling simple metals like Nb and Pb, it also
provides a convincing explanation of the superconducting features of more complex materials 
such as MgB$_2$~\cite{louie}.

In spite of its tremendous success, Eliashberg theory, in its practical implementation, has to
be considered a semi-phenomenological theory. While the electron-phonon interaction is 
perfectly accounted for, correlation effects due to the electron-electron Coulomb repulsion 
are difficult to handle in this theory. Those effects are condensed in a single parameter,
$\mu^*$, which represents a measure of the effective electronic repulsion. Although $\mu^*$
could, in principle, be calculated by diagrammatic techniques~\cite{scalapino}, first principles 
estimates of $\mu^*$ are extremely hard to make, and in practice, $\mu^*$ is treated as an
adjustable parameter, usually chosen such that the experimental T$_\text{c}$ is reproduced. 

The goal of this work is to develop a true ab-initio theory for superconductivity which 
does not contain any adjustable parameters. The crucial point is to treat the electron-phonon
interaction and the Coulombic electron-electron repulsion on the same footing. This is 
achieved within a density-functional framework.

Density functional theory (DFT)\cite{HK,KS,Grossbook} enjoys enormous popularity as an
electronic-structure method in solid-state physics, quantum chemistry and materials science.
DFT combines good accuracy with moderate numerical effort  and is often the method of choice
especially for large molecules and solids with a big unit cell. DFT is based on the 
Hohenberg-Kohn~\cite{HK} theorem which ensures a rigorous 1--1 correspondence 
between the ground-state density and the external potential. At finite temperature,
the correspondence holds~\cite{mermin} between the density in thermal 
equilibrium and the external potential. As a consequence, 
all physical observables of an interacting electron system become
functionals of the density. The practical implementation of DFT rests on the 
Kohn-Sham~\cite{KS} scheme which maps the interacting system of interest on an 
auxiliary non-interacting system with the same ground-state thermal density.

Traditional DFT, by its very nature, inevitably involves the Born-Oppenheimer 
approximation: One is supposed to calculate the electronic ground-state/thermal 
density that corresponds 1--1 to the electrostatic potential of clamped nuclei.
To overcome this limitation Kreibich and Gross (KG)~\cite{kreibich} recently presented 
a multicomponent DFT which treats both electrons and nuclei quantum mechanically on the
same footing. The KG theory involves two ``densities'': The electronic density
$n(\br)$ referring to a body-fixed coordinate frame, and the diagonal $\Gamma(\underline{\bR})$
of the nuclear $N$-body density matrix. The exchange-correlation functional 
appearing in the resulting Kohn-Sham equations depends on both ``densities'' and
contains, formally, all non-adiabatic couplings between the electrons and the nuclei.

Hence, in principle, the KG framework should be able to describe (conventional) 
superconductivity. In practice, however, it is not advisable to attempt such a
description. The situation is quite similar to the DFT treatment of magnetic 
effects: By virtue of the Hohenberg-Kohn theorem, magnetic effects 
can be described on the basis of the density alone. In particular, the order 
parameter of spin magnetism, the spin magnetization $\bem(\br)$, is a functional of the 
density $\bem = \bem[n]$. However, this functional has to be highly non-local 
and its explicit form is unknown. Therefore, it is advisable to include the order 
parameter $\bem(\br)$ as an additional ``density'' in the DFT formulation~\cite{barth}.
This version of DFT is known as spin-DFT. It is the standard form of DFT which is
employed in all practical applications.

A similar idea was suggested in 1988 by Oliveira, Gross and Kohn (OGK)~\cite{OGK} to 
treat superconductors in a DFT framework. OGK proposed the inclusion of the order parameter
$\chi(\br,\br')$ that characterizes the superconducting phase as a basic ``density'' in the
DFT formulation. OGK dealt with singlet order parameters. A generalization to triplet 
order parameters was given later~\cite{capelle}.

The complexities of the many-body 
problem were cast into an exchange and correlation term, but in contrast to ordinary density 
functional theory where a variety of functionals has appeared over the past thirty years,
very few exchange-correlation functionals have been proposed for the superconducting state. 
To our knowledge, only a local density approximation
describing the purely electronic interactions has been presented\cite{SC-LDA}. However,
the usefulness of the OGK approach was demonstrated by Gy\"orffy and coworkers in their study of
niobium and YBa$_2$Cu$_3$O$_7$ using a semi-phenomenological parametrization of the exchange-correlation
functional\cite{balazs,gyorffy:98}. 

The OGK formulation was triggered by the discovery of high-$T_\text{c}$ superconductors~\cite{hightc}
where an electronic pairing mechanism was believed to be dominant. Hence, the 
OGK description treats the Coulomb repulsion between electrons formally exactly while the 
electron phonon coupling only enters through a given, non-retarded BCS-type 
electron-electron interaction, i.e. strong electron-phonon coupling cannot be
dealt with.

In this work we develop a DFT for superconductors based on the three densities
$n(\br)$, $\chi(\br,\br')$ and $\Gamma(\underline{\bR})$. The formalism can thus be viewed as a
superconducting generalization of KG or as a strong coupling generalization
of OGK. It leads to a set of  - formally exact - Kohn-Sham equations
for electrons and nuclei. These equations contain exchange-correlation
potentials which are universal functionals of the three densities
$n, \chi$ and $\Gamma$. For the time being, we do not study
effects found in the presence of magnetic fields. Those can be accomodated by 
generalizing the framework to include the current density as an additional 
variable \cite{Paris,Wacker}.

The success of any density functional theory
crucially depends on the availability of accurate approximations for
the exchange-correlation functionals. The main body of this article
is devoted to the construction of such approximate exchange-correlation
functionals. Diagrammatic many-body perturbation theory is used for
this purpose. In a second article (henceforth referred to as II),
these approximate functionals are employed to calculate superconducting
properties of elemental metals. 

The present paper is organized as follows: In Sect.~\ref{sec:2} we derive a multicomponent DFT for 
the superconducting state. This theory leads to a set of Kohn-Sham equations that 
are described in the following section. Sect.~\ref{sec:4} is devoted to the development
of Kohn-Sham perturbation theory. The resulting exchange-correlation potentials are discussed
in Sect.~\ref{sec:5}. A simple BCS-like model is described in Sect.~\ref{sec:6}.
This model is used, in Sect.~\ref{sec:7}, to analyze the approximate exchange-correlation
kernels entering the linearized DFT gap equation. Finally, in Sect.~\ref{sec:8}, the
exchange-correlation contributions to the non-linear gap equation are discussed.

\section{Multicomponent DFT for superconductors}
\label{sec:2}

It is clear that a balanced treatment of the electron-phonon and Coulomb 
interactions has to start from the many-body electron-nuclear Hamiltonian
(atomic units are used throughout this paper)
\be
  \label{Hamiltonian}
  \hat{H} = \hat{T}^\text{e} + \hat{U}^\text{ee} + 
  \hat{T}^\text{n} + \hat{U}^\text{nn}
  + \hat{U}^\text{en}
  \,,
\ee
where \(\hat{T}^\text{e}\) represents the electronic kinetic energy,
\(\hat{U}^\text{ee}\) the electron-electron interaction,
\(\hat{T}^\text{n}\) the nuclear kinetic energy and
\(\hat{U}^\text{nn}\) the Coulomb repulsion between the nuclei. 
The interaction between the electrons and the nuclei is
described by the term:
\be
  \hat{U}^\text{en} = -  \sum_\sigma \mdint{r}{R}
  \hat{\Psi}^\dagger_\sigma(\br) \hat{\Phi}^\dagger(\bR)
  \frac{Z}{|\br - \bR|}\hat{\Phi} (\bR) \hat{\Psi}_\sigma(\br)
  \,,
\ee
where $\hat{\Psi}^\dagger_\sigma(\br)$ and $\hat{\Phi}^\dagger(\bR)$
are respectively electron and nuclear creation operators. (For simplicity
we assume the nuclei to be identical, and we neglect the nuclear spin
degrees of freedom. The extension of this framework to a more general
case is straightforward.) Note that there is no external potential in
the Hamiltonian.

To develop a multicomponent DFT for the electron-nuclear system we
have to proceed with care. The Hamiltonian~(\ref{Hamiltonian})
describes a translationally invariant and isotropic system. Thus, both
the electronic and nuclear one-particle densities are constant and
therefore not useful to characterize the system. This problem can be
solved by adopting a body-fixed reference
frame~\cite{kreibich,pLeeuwen}.  In this article we are interested in
infinite solids where the nuclei perform small oscillations around the
equilibrium positions. Furthermore, we assume that the solid is not
rotating as a whole. Fortunately, in this case, the body-fixed
reference frame coincides with the normal Cartesian system commonly
used to describe solids. The situation is very different for finite
systems, which have to be handled by using appropriate internal
coordinates.

In order to formulate a Hohenberg-Kohn type statement, the Hamiltonian~(\ref{Hamiltonian})
is generalized to
\be
  \label{eq:total_hamiltonian}
  \op{H} = \hat{T}^\text{e} + \hat{T}^\text{n} + \hat{U}^\text{en} + \hat{U}^\text{ee}+
  \hat{V}_\text{ext}^\text{e} + \hat{V}_\text{ext}^\text{n} + \op{\Delta}_\text{ext} -\mu\op{N}
  \,.
\ee
The external potential for the electrons is defined as
\be
  \hat{V}_\text{ext}^\text{e} = \sum_{\sigma} \mint{r} \hat{\Psi}^\dagger_\sigma(\br)
  v^\text{e}_{\rm ext}(\br) \hat{\Psi}_\sigma(\br)
  \,. 
\ee
Since, at this level, the nuclei are taken into account explicitly,
the lattice potential is not treated as an external field,
but is included via the interaction term $\hat{U}^\text{en}$.
The term $\hat{V}_\text{ext}^\text{e}$ is introduced as a mathematical trick
to prove the Hohenberg-Kohn theorem, and will be taken to zero at the end
of the derivation. $\hat{V}_\text{ext}^\text{n}$ is a multiplicative $N$-body
operator with respect to the nuclear coordinates
\be
  \hat{V}_\text{ext}^\text{n} = \mint{\underline{R}} v^\text{n}_{\rm ext}
  (\underline{\bR}) \op{\Gamma}( \underline{\bR} )
  \,,
\ee
where we have defined $\underline{\bR} = \{\bR_1,\bR_2,\cdots,\bR_N\}$, 
$\D^3\underline{R} = \D^3 R_1 \D^3 R_2 \cdots \D^3 R_N$, and
\be
  \hat\Gamma(\underline{\bR}) = \hat\Phi^\dagger(\bR_1) \dots \hat\Phi^\dagger(\bR_N)
  \hat\Phi(\bR_N) \dots \hat\Phi(\bR_1)
\ee
is the diagonal part of the nuclear $N$-particle density matrix operator.
Note that the term $\hat{V}_\text{ext}^\text{n}$ includes the interaction between the nuclei 
$\hat U^\text{nn}$ (to which it reduces if no other external nuclear potentials are
present). The term
\be
  \label{eq:anomalousextpot}
  \hat{\Delta}_\text{ext} =  - \mdint{r}{r'} \left[ \Delta^*_{\rm ext}(\br,\br') 
  \hat{\Psi}_\up(\br) \hat{\Psi}_\down(\br')  + h.c. \right]
\ee
describes an external pairing field, and usually vanishes unless our system is in the
proximity of an adjacent superconductor. However, this term is required
to break the gauge invariance of the Hamiltonian, and the limit 
$\Delta_{\rm ext}\to 0$ can only be taken at the end of the derivation.
Note that the term~(\ref{eq:anomalousextpot}) describes a singlet pairing field.
The extension to triplet superconductors is straightforward \cite{capelle}.
Finally, $\mu$ stands for the chemical potential, and $\hat N$
is the number operator for the electrons (we treat the electronic
degrees of freedom in a grand-canonical ensemble).

Our multicomponent formulation is based on three densities:

i) The electronic density \be n(\br) = \sum_\sigma \langle \op{\Psi}_\sigma^\dagger(\br)
\op{\Psi}_\sigma(\br) \rangle \,, \ee which is defined in the usual way.  The
bracket $\langle\cdots\rangle$ denotes the thermal average
$\langle\op{A}\rangle=\Tr{\op{\rho}_0\op{A}}$, with the grand canonical statistical
density operator $\op{\rho}_0 = {\E^{-\beta\op{H}}} / {\Tr{\E^{-\beta\op{H}}}}$
in the superconducting state. We furthermore define the inverse
temperature $\beta = 1/T$.

ii) The anomalous density
\be
  \chi(\br,\br') = \langle\hat{\Psi}_\up(\br) \hat{\Psi}_\down(\br') \rangle
  \,,
\ee
which is the order parameter characterizing the singlet superconducting state. This quantity 
is finite for superconductors below the transition temperature and zero above this temperature.

iii) To describe the nuclear degrees of freedom,
we use the diagonal part of the nuclear $N$-particle density matrix 
$\Gamma( \underline{\bR} ) = \langle \op\Gamma(\underline{\bR}) \rangle$.
Alternatively, one could define a multicomponent DFT using the one-particle density for the nuclei
$n_{\rm n}(\bR) = \langle \op{\Phi}^\dagger(\bR) \op{\Phi}(\bR) \rangle$. However, in the
following it will be convenient to transform the nuclear degrees of freedom to collective 
(phonon) coordinates. Using $n_{\rm n}(\bR)$ would lead to a {\it one body} equation for
non-interacting nuclei. Thus, the nuclear Kohn-Sham equation would not lead to realistic
phonons with a proper dispersion relation. Only Einstein phonons could be present in
this system. This is also clear from the fact that a system of non-interacting particles
does not exhibit collective modes. With our choice
of $\Gamma(\underline{\bR})$, the nuclei obey a $N$-particle equation, very similar to the
Born-Oppenheimer equation, and where phonon coordinates can be easily introduced.

As usual, the Hohenberg-Kohn theorem guarantees a one-to-one mapping
between the set of the densities  $\{n(\br),\chi(\br,\br'),\Gamma(\underline{\bR})\}$
in thermal equilibrium and the set of their conjugate potentials 
$\{v_{\rm ext}^\text{e}(\br),\Delta_{\rm ext}(\br,\br'),v_{\rm ext}^\text{n}(\underline{\bR})\}$. 
As a consequence all observables are functionals of the
set of densities. Finally it assures that the grand canonical potential,
\begin{multline}
  \label{eq:intomega}
  \Omega[n,\chi,\Gamma] = F[n,\chi,\Gamma] + \mint{r} n(\br) [v^\text{e}_{\rm ext}(\br) - \mu]
  \\ - \mdint{r}{r'} \left[ \chi(\br,\br') \Delta_{\rm ext}^*(\br,\br') + h.c. \right]
  \\ + \mint{\underline R} \, \Gamma(\underline{\bR}) v_{\rm ext}^\text{n}(\underline{\bR})
  \,,
\end{multline}
is minimized by the equilibrium densities. We use the notation $A[f]$ to denote that
$A$ is a functional of $f$. The functional $F[n,\chi,\Gamma]$ is
universal, in the sense that it does not depend on the external potentials, and
is defined by
\begin{multline}
  \label{eq:intF}
  F[n,\chi,\Gamma] = T^\text{e}[n,\chi,\Gamma] + T^\text{n}[n,\chi,\Gamma] \\
   + U^\text{en}[n,\chi,\Gamma] + U^\text{ee}[n,\chi,\Gamma]
  - \frac{1}{\beta} S[n,\chi,\Gamma]
\end{multline}
where $S$ stands for the entropy of the system
\be
  S[n,\chi,\Gamma] = -\Tr{\op\rho_0[n,\chi,\Gamma] \log(\op\rho_0[n,\chi,\Gamma])}
  \,.
\ee
The proof of the theorem follows closely the proof of the Hohenberg-Kohn theorem 
at finite temperatures \cite{mermin} and will not be presented here.

In standard DFT one normally defines a Kohn Sham system, i.e. a non-interacting system
chosen such that it has the same ground-state density as the interacting one. In our formulation, the 
Kohn-Sham system consists of non-interacting (superconducting) electrons, and 
{\it interacting} nuclei. It is described by the thermodynamic potential 
[cf. Eq.~(\ref{eq:intomega})]
\begin{multline}
  \label{eq:nonintomega}
  \Omega_\KS[n,\chi,\Gamma] = F_\KS[n,\chi,\Gamma] + \mint{r} n(\br) [v^\text{e}_\KS(\br) - \mu_\KS]
  \\ - \mdint{r}{r'} \left[ \chi(\br,\br') \Delta_\KS^*(\br,\br') + h.c. \right]
  \\ + \mint{\underline R} \, \Gamma(\underline{\bR}) v_\KS^\text{n}(\underline{\bR})
  \,,
\end{multline}
where $F_\KS$ if the counterpart of~(\ref{eq:intF}) for the Kohn-Sham system, i.e.,
\be
F_\KS[n,\chi,\Gamma] = T_\KS^\text{e}[n,\chi,\Gamma] + T_\KS^\text{n}[n,\chi,\Gamma]  - \frac{1}{\beta} S_\KS[n,\chi,\Gamma] \,.
\ee
Here $T_\KS^\text{e}[n,\chi,\Gamma],T_\KS^\text{n}[n,\chi,\Gamma]$ and $S_\KS[n,\chi,\Gamma]$
are the electronic and nuclear kinetic energies and the entropy of the
Kohn-Sham system, respectively.

From Eq.~(\ref{eq:nonintomega}) it is clear that the Kohn-Sham nuclei
interact with each other through the $N$-body potential
$v_\KS^\text{n}(\underline{\bR})$ while they do not interact with the
electrons.

By applying the Hohenberg-Kohn theorem to both the interacting and the
non-interacting systems, and requiring the densities of the Kohn-Sham system
to reproduce the densities of the fully interacting one, we can
identify the expressions for the effective Kohn-Sham potentials.  As
usual, these include contributions from external fields, Hartree,
and exchange-correlation terms. The latter account for all the
many-body effects stemming from the electron-electron and
electron-nuclear interactions. To simplify the expressions, we now set
the auxiliary external potentials to zero.

The Kohn-Sham potential for the electrons $v^\text{e}_\KS(\br)$ reads as
\begin{multline}
  \label{eq:vksmulti}
  v^\text{e}_\KS[n,\chi,\Gamma] (\br) = - Z \sum_\alpha \mint{\underline{\bR}} 
  \frac{\Gamma(\underline{R})}{|\br - \bR_\alpha |} \\
  + \mint{r'} \frac{n(\br')} {|\br - \br'|}
  + v^\text{e}_{\rm xc}[n,\chi,\Gamma](\br)
  \,.
\end{multline}
The first term, the electron-nuclear Hartree potential, reduces to the
usual nuclear attraction potential if we assume that the nuclei are
classical and perfectly localized at their equilibrium positions. This
term is usually treated as the external potential in standard DFT. The
last two contributions to $v^\text{e}_\KS(\br)$ are respectively the
Hartree potential, which accounts for the classical repulsion between
the electrons, and the exchange-correlation term.

The anomalous Kohn-Sham potential $\Delta_\KS$ is given by
\be
  \label{eq:DeltaKS}
  \Delta_\KS[n,\chi,\Gamma](\br, \br') = -\frac{\chi(\br, \br')}{|\br-\br'|}
  +\Delta_{\rm xc}[n,\chi,\Gamma](\br, \br') \,.
\ee
Note that the first term, the so-called anomalous Hartree potential,
gives rise to a positive contribution to the energy.

Finally, the nuclear potential is 
\begin{multline}
  \label{eq:vKSnuclei}
  v^\text{n}_\KS[n,\chi,\Gamma] (\underline{\bR}) = \sum_{\alpha \neq \beta} 
  \frac{Z^2}{|\bR_\alpha - \bR_\beta|} 
  \\ - Z \sum_\alpha \mint{r} \frac{n(\br)}{|\br - \bR_\alpha|}
  +v^\text{n}_{\rm xc}[n,\chi,\Gamma](\underline{\bR})\,.
\end{multline}
The first term stems from $\op{U}^\text{nn}$, and describes the bare nuclear-nuclear repulsion. 
The second is the nuclear-electron Hartree term and is the counterpart of the first term in 
Eq.~(\ref{eq:vksmulti}).

As in standard DFT, the exchange-correlation potentials are defined as functional derivatives
\begin{subequations}
\label{eq:xc_potentials}
\bea
  v^\text{e}_{\rm xc}[n,\chi,\Gamma](\br) & = & \phantom{-}
  \frac{\delta F_{\rm xc}[n, \chi, \Gamma]}{\delta n(\br)} \,,
  \\
  \label{eq:xc_potentials_anomalous}
  \Delta_{\rm xc}[n,\chi,\Gamma](\br,\br') & = &
  -\frac{\delta F_{\rm xc}[n, \chi, \Gamma]}{\delta \chi^*(\br,\br')}\,,
  \\
  v^\text{n}_{\rm xc}[n,\chi,\Gamma](\underline{\bR}) & = & \phantom{-}
  \frac{\delta F_{\rm xc}[n, \chi, \Gamma]}{\delta \Gamma(\underline{\bR})}
  \,.
\eea
\end{subequations}
The exchange-correlation free energy is defined through the equation
\begin{multline}
  F[n, \chi, \Gamma] = F_\KS[n, \chi, \Gamma]+ F_{\rm xc}[n, \chi, \Gamma]
  \\ +U^\text{nn}[\Gamma] +E^\text{ee}_{\rm H}[n,\chi] + E^\text{en}_{\rm H}[n,\Gamma]
  \,.
\end{multline}
There are two contributions to $E^\text{ee}_{\rm H}$, one stemming from the electronic
Hartree potential, and the other from the anomalous Hartree potential
\begin{multline}
  \label{eq:el_hartree}
  E^\text{ee}_{\rm H}[n,\chi] = 
  \frac{1}{2} \mdint{r}{r'} \frac{n(\br) n(\br')}{|\br-\br'|} \\
  + \mdint{r}{r'} \frac{|\chi(\br,\br')|^2}{|\br-\br'|}
  \,.
\end{multline}
Finally, $E^\text{en}_{\rm H}$ denotes the electron-nuclear Hartree energy
\be
  E^\text{en}_{\rm H}[n,\Gamma] = - Z \sum_\alpha \mdint{r}{\underline R} 
  \frac{n(\br) \Gamma(\underline{\bR}) }{|\br - \bR_\alpha|}
  \,.
\ee

\section{The Kohn-Sham equations}
\label{sec:3}

The problem of minimizing the Kohn-Sham grand canonical potential
(\ref{eq:nonintomega}) can be transformed into a set of three
differential equations that have to be solved self-consistently: Two
of them are coupled and describe the electronic degrees of freedom.
Their algebraic structure is similar to the Bogoliubov-de Gennes
equations. The third is an equation for the nuclei resembling the
familiar nuclear Born-Oppenheimer equation.

\subsection{Electronic equations}

The Kohn-Sham Bogoliubov-de Gennes (KS-BdG) equations~\cite{bogoliubov} read
\begin{subequations}
\label{KS-BdG}
\bea
\left[ -\frac{\nabla^2}{2} + v^\text{e}_\KS(\br) - \mu \right] u_i(\br)
&+& \mint{r'} \Delta_\KS(\br,\br') v_i(\br') \nonumber \\ 
&=& \tilde{E}_i \, u_i(\br) \,,\\
- \left[ -\frac{\nabla^2}{2} + v^\text{e}_\KS(\br) - \mu \right] v_i(\br)
&+& \mint{r'} \Delta^*_\KS(\br,\br') u_i(\br') \nonumber \\
&=& \tilde{E}_i \, v_i(\br)\,, 
\eea
\end{subequations}
where $u_i(\br)$ and $v_i(\br)$ are the particle and hole amplitudes.
This equation is very similar to the Kohn-Sham equations in the OGK
formalism~\cite{OGK}.  However, in our formulation the lattice
potential is not considered as an external potential but enters via
the electron-ion Hartree term. Furthermore, our exchange-correlation
potentials depend parametrically on the nuclear density matrix, and
therefore on the phonons.

Although these equations have the structure of static equations,
they contain, in principle, all retardation effects 
through the exchange-correlation potentials.

A direct solution of the Kohn-Sham Bogoliubov-de Gennes
equations\cite{balazs} is faced with the problem that one needs
extremely high accuracy to resolve the superconducting energy gap,
which is about three orders of magnitude smaller than typical
electronic energies.  At the same time, one has to cover the whole
energy range of the electronic band structure.  The so-called
decoupling approximation~\cite{GrossKurth,dissKurth,dissLueders} relieves the 
problem by
separating these two energy scales.

The particle and hole amplitudes can be expanded in the complete set
of wave functions $\{\varphi_i\}$ of the normal-state Kohn-Sham equation
\be
  \label{eq:normalKS}
  \left[- \frac{\nabla^2}{2} + v^\text{e}_\KS[n,\chi,\Gamma](\br) - \mu \right] \varphi_i(\br) =
  \epsilon_i \, \varphi_i(\br)
\ee
which can be solved by standard band-structure methods. 
The decoupling approximation then implies the following form for the
particle and hole amplitudes, where only one term of the expansion
is retained: 
\be
  \label{eq:decapprox}
  u_i(\br) \approx u_i \varphi_i(\br)
  \quad ; \quad 
  v_i(\br) \approx v_i \varphi_i(\br)
  \,.
\ee
In this way the eigenvalues in Eq.~(\ref{KS-BdG}) become $\tilde E_i =
\pm E_i$, where 
\be
E_i = \sqrt{\xi_i^2+|\Delta_i|^2} \,,
\ee
and $\xi_i = \epsilon_i-\mu$. This form of the eigenenergies allows the
interpretation of the matrix elements of the pair potential as the gap
function of the superconductor.  The coefficients $u_i$ and $v_i$ also
have simple expressions within this approximation
\begin{subequations}
\bea
  u_i & = & \frac{1}{\sqrt{2}}{\rm sgn}(\tilde E_i) \E^{\I\phi_i}
  \sqrt{1+\frac{\xi_i}{\tilde E_i}} \,,
  \\
  v_i & = & \frac{1}{\sqrt{2}} \sqrt{1-\frac{\xi_i}{\tilde E_i}}
  \,.
\eea
\end{subequations}
Finally, the matrix elements $\Delta_i$ are defined as
\be
  \label{eq:delta_i}
  \Delta_i = \mdint{r}{r'} \varphi^*_i(\br)\Delta_\KS(\br,\br')\varphi_i(\br')
  \,,
\ee
and $\phi_i$ is the phase $\E^{\I\phi_i} = \Delta_i/|\Delta_i|$. 
Within the decoupling approximation, the normal and the anomalous densities can be easily obtained 
from the particle and hole amplitudes
\begin{subequations}
\label{eq:el-densities}
\bea
  n(\br) & = & \sum_i \left[1-\frac{\xi_i}{E_i}
  \tanh\left(\frac{\beta}{2}E_i\right)\right]
  |\varphi_i(\br)|^2
  \\
  \chi(\br,\br') & = & \frac{1}{2}\sum_i
  \frac{\Delta_i}{E_i} \tanh\left(\frac{\beta}{2}E_i\right)
  \varphi_i(\br)\varphi^*_i(\br')
  \,.
\eea
\end{subequations}

The validity and limitations of the decoupling approximation will be discussed in 
detail in II.

\subsection{Nuclear equation}

The Kohn-Sham equation for the nuclei has the form
\be
  \label{KSNa}
  \left[-\sum_{\alpha}\frac{\nabla^2_\alpha}{2M}+
  v_{\KS}^\text{n}(\underline{\bR})\right] \Phi_l(\underline{\bR})
  = {\cal E}_l \Phi_l(\underline{\bR})
  \,.
\ee
This equation has the same structure as the usual nuclear Born-Oppenheimer
equation. We emphasize that the Kohn-Sham equation~(\ref{KSNa}) does not rely on
any approximation and is, in principle, exact. As already mentioned, we are
interested in solids at relatively low temperature, where the nuclei perform
small amplitude oscillations around their equilibrium positions. In this case,
we can expand the $v_{\KS}^\text{n}[n,\chi,\Gamma]$ in a Taylor series around the equilibrium 
positions, and transform the nuclear degrees of freedom into collective 
(phonon) coordinates. In harmonic order, the nuclear Kohn-Sham Hamiltonian
then reads
\be
  \label{eq:h_phonon}
  \op{H}_\KS^{\rm ph} = \sum_{\lambda,\bq} \Omega_{\lambda,\bq}
  \left[\op{b}^\dagger_{\lambda,\bq}\op{b}{\dagger}_{\lambda,\bq} + \frac{1}{2}\right]
  \,,
\ee
where $\Omega_{\lambda,\bq}$ are the phonon eigenfrequencies, and $\op{b}_{\lambda,\bq}$
destroys a phonon from the branch $\lambda$ and wave-vector $\bq$. Note that
the phonon eigenfrequencies are functionals of the set of densities 
$\{n,\chi,\Gamma\}$, and can therefore be affected by the superconducting 
order parameter.
Within the harmonic approximation, the nuclear density matrix $\Gamma(\underline{\bR})$
is given by
\be
\label{eq:G_phonon}
\Gamma(\underline{\bR}) = \sum_{\lambda,\bq} n_\beta(\Omega_{\lambda,\bq}) | h_{\lambda,\bq}(\bQ) |^2
\ee
where $n_\beta(\Omega)$ denote the Bose occupation numbers and $h_{\lambda,\bq}(\bQ)$ 
are harmonic oscillator wave functions referring to the collective coordinates $\bQ$.

\subsection{Gap equation}
\label{seq:gap_eq}

\begin{figure}[Htb]
\includegraphics{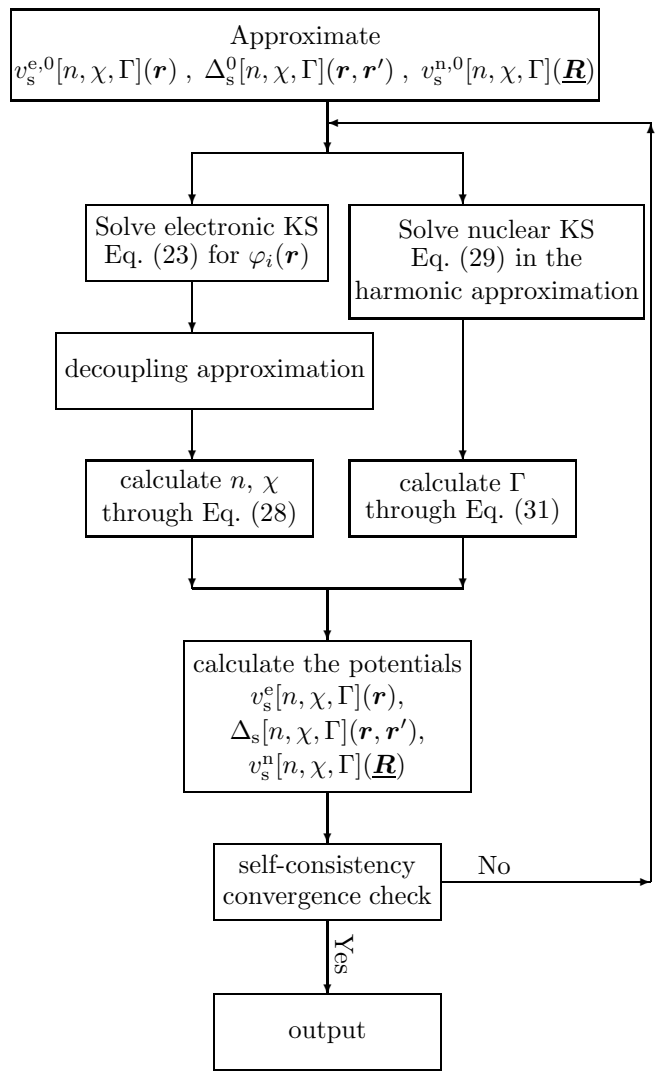}
\caption{\label{fig:cycle1}
Schematic flow chart for the iterative Kohn-Sham scheme within the decoupling 
approximation.}
\end{figure}

In Fig.~\ref{fig:cycle1} we sketch the Kohn-Sham self-consistent procedure
within the decoupling approximation. We start by finding suitable approximations
for the Kohn-Sham potentials to start the cycle: for $v^{\text{e},0}_\KS[n,\chi,\Gamma]$ we can 
use the Kohn-Sham potential stemming from a standard DFT calculation for the 
non-superconducting ground-state, i.e, $v^\text{GS}_\KS[n]$. 
This is a very good approximation as the dependence of
$v^\text{e}_\KS[n,\chi,\Gamma]$ on $\chi$ and $\Gamma$ is certainly very weak for the usual
superconductors at low temperature; The starting pair potential $\Delta^0_\KS[n,\chi,\Gamma]$
can be approximated by a square well centered at the Fermi energy with width of the order of the Debye
frequency and height computed from a BCS model; Finally, for $v^{\text{n},0}_\KS[n,\chi,\Gamma]$ 
we can use the ground-state Born-Oppenheimer potential. The next two steps of the self-consistent
cycle can be performed in parallel:
\begin{itemize}
\item Equation~(\ref{eq:normalKS}) is solved to obtain the wave-functions $\varphi_i$'s and the 
eigenenergies $\epsilon_i$'s. These can then be used within the decoupling approximation, 
equation~(\ref{eq:decapprox}), to calculate the normal and anomalous densities through 
Eqs.~(\ref{eq:el-densities}). We note that the chemical potential $\mu$ 
entering Eq.~(\ref{eq:normalKS}) has to be adjusted at every iteration, such that the
density $n(\br)$ integrates to the correct particle number $N$.

\item With $v^\text{n}_\KS[n,\chi,\Gamma]$ we solve the nuclear equation~(\ref{KSNa}) by expanding
the nuclear potential to harmonic order to obtain the phonon eigenfrequencies
and eigenmodes. The nuclear density matrix $\Gamma$ then follows from Eq.~(\ref{eq:G_phonon}).
\end{itemize}
Finally, the set of densities $\{n,\chi,\Gamma\}$ is used to evaluate the new Kohn-Sham
potentials $\{v_\KS^\text{e}, \Delta_\KS, v^\text{n}_\KS\}$ from the 
definitions~(\ref{eq:vksmulti},\ref{eq:DeltaKS},\ref{eq:vKSnuclei}). At this point, if
self-consistency is reached, the cycle is stopped. Otherwise, the new potentials are used
to restart the cycle.

It is clear that, even within the decoupling approximation, the task of solving the
self-consistent cycle depicted in Fig.~\ref{fig:cycle1} is rather demanding. Furthermore,
we are required to provide (good) approximations for the three functionals
$v^\text{e}_\KS[n,\chi,\Gamma]$, $\Delta_\KS[n,\chi,\Gamma]$, and $v^\text{n}_\KS[n,\chi,\Gamma]$.
As our objective is to study superconductivity, we will make two simplifying assumptions:
i) $v^\text{e}_\KS[n,\chi,\Gamma]$ can be well approximated by the ground state functional
used in standard density functional theory $v^\text{GS}_\KS[n]$. As the energy scale of the
phonons and of the superconducting gap is three orders of magnitude smaller than
electronic energy scales (like the Fermi energy) this is a very reasonable assumption.
ii) The nuclear functional $v^\text{n}_\KS[n,\chi,\Gamma]$ is approximated by the ground-state
Born-Oppenheimer energy surface.  It is well known that calculations 
performed within the Born-Oppenheimer approximation yield phonon frequencies that are in 
excellent agreement with experimental dispersions\cite{bofre}. We therefore expect this to be
an excellent approximation to the Kohn-Sham phonons. However, we are neglecting
the effect of the superconducting pair potential on the phonon dispersion. This
effect has been measured experimentally~\cite{pCardona}, and it turns out to be quite small.
Note that these two approximations are equivalent to fixing $v^\text{e}_\KS[n,\chi,\Gamma]$ and
$v^\text{n}_\KS[n,\chi,\Gamma]$ to $v^{\text{e},0}_\KS[n,\chi,\Gamma]$ and 
$v^{\text{n},0}_\KS[n,\chi,\Gamma]$.

By inserting Eqs.~(\ref{eq:el-densities}) in Eq.~(\ref{eq:DeltaKS}) and by subsequently inserting
Eq.~(\ref{eq:DeltaKS}) on the right-hand side of Eq.~(\ref{eq:delta_i}), we obtain the
DFT gap equation
\be
  \label{eq:nlgap}
  \Delta_i = \Delta_{{\rm Hxc}\;i}[\mu, \Delta_i]
  \,,
\ee
where $\Delta_{\rm Hxc}$ stands for the sum of the Hartree and exchange-correlation
contributions to the functional.
Note that through the exchange-correlation functional the
right-hand side of Eq.~(\ref{eq:nlgap}) becomes a highly complicated functional
of $\mu$ and $\Delta_i$ . The dependence on the gap function is totally different 
from the usual mean-field  gap equation (cf. Sect~\ref{sec:8}).

After these simplifying approximations, a Kohn-Sham calculation proceeds as follows:
i)~Using a standard band structure code, the ground state wave-functions and band
structure are obtained. ii)~The Born-Oppenheimer phonon frequencies and eigenmodes 
are obtained from linear-response\cite{bofre} calculations, again using
standard tools widely available to the community. iii)~The gap equation~(\ref{eq:nlgap})
is iterated until self-consistency is achieved. We can now see how the different energy 
scales are separated: The normal density, the anomalous density, and the phonon properties
are obtained from three separate equations.

In the vicinity of the transition temperature, the anomalous density will be vanishingly
small. In this regime, the gap-equation can be linearized in $\chi$, leading to
\be
  \label{lin-gap}
  \breve \Delta_i = - \frac{1}{2} \sum_{j} {\cal F}_{{\rm Hxc}\;i,j}[\mu]
  \frac{\tanh\left(\frac{\beta}{2}\xi_j\right)}{\xi_j} \breve \Delta_j \,, \
\ee
where the anomalous Hartree exchange-correlation kernel of the
homogeneous integral equation reads 
\be
  \label{eq:deffxc}
  {\cal F}_{{\rm Hxc}\;i,j}[\mu] = - \left. \frac{\delta \Delta_{{\rm Hxc} \, i}}{\delta \chi_{j}}\right|_{\chi = 0}
  = \left. \frac{\delta^2 (E^\text{ee}_{\rm H} + F_{\rm xc})}
  {\delta \chi^*_i \delta \chi_j} \right|_{\chi = 0}
  \,.
\ee
We emphasize that the linearized gap equation can only be used to calculate the superconducting
transition temperature. In particular, the function $\breve \Delta_i$ that stems from the solution
of this equation does not have any physical interpretation.

We can gain further insight into Eq.~(\ref{lin-gap}) if we separate the kernel 
${\cal F}_{{\rm Hxc}\;i,j}$ into a purely diagonal term, ${\cal Z}_i$, and a non-diagonal part, ${\cal K}_{i,j}$
\be
  \label{eq:gap} 
  \breve \Delta_i = - {\cal Z}_i[\mu] \;\breve\Delta_i -\frac{1}{2}\sum_j
  {\cal K}_{i,j}[\mu] 
  \frac{\tanh\left(\frac{\beta}{2}\xi_j\right)}{\xi_j} 
  \breve \Delta_j
  \,,
\ee
This equation has the same structure as the BCS gap equation with the
kernel ${\cal K}_{i,j}$ replacing the model interaction of BCS theory.
On the other hand, ${\cal Z}_i$ plays a similar role as the
renormalization term in the Eliashberg equations.  This similarity
allows one to interpret the kernel ${\cal K}_{i,j}$ as an effective
interaction responsible for the binding of the Cooper pairs.  The
function ${\cal K}_{i,j}$ is a quantity of central importance in the
density functional theory for superconductors. By studying ${\cal K}_{i,j}$ for a specific 
material as a function of $i$ and $j$ one
can learn which orbitals are responsible for superconductivity and, ultimately,
by identifying those parts of the exchange-correlation functional (phononic/Coulombic) that
lead to an effective attraction, one can trace the mechanism responsible
for the superconducting state.

Note that Eq.~(\ref{eq:gap}) is considerably simpler than the
Eliashberg equations as it does not dependent explicitly on the
frequency.  However, phonon retardation effects are included through
the exchange-correlation terms. Furthermore, Eq.~(\ref{eq:gap}) is not
a mean-field equation like in BCS theory but contains correlation
effects. A linearized gap equation can also be derived without the
decoupling approximation~\cite{LuedersGross:95,dissLueders}, leading
to a similar equation, but with a four-point kernel.  From this point
of view, the decoupling approximation can be viewed as a diagonal
approximation to this four-point kernel, neglecting the corresponding
off-diagonal elements.

\section{Kohn-Sham perturbation theory}
\label{sec:4}
In the previous sections we showed how to develop a density functional
theory for the superconducting state. The main equation of this
theory, the gap equation~(\ref{eq:nlgap}), allows, in principle, the
calculation of the superconducting gap for any system. However, to solve
this equation one needs approximations for $\Delta_{\rm xc}$, the
exchange-correlation contribution to the Kohn-Sham pair potential. In
the following, we will develop such approximations by applying Kohn-Sham
perturbation theory, as described by G\"orling and Levy~\cite{GoerlingLevy}, to 
superconducting systems~\cite{dissLueders,dissMarques}.  This perturbation
theory, that will treat both the electron-electron and electron-phonon
interactions on the same footing, is a generalization of the method
used by Kurth {\it et al.} to calculate the exchange-correlation
energy of the uniform superconducting electron gas~\cite{SC-LDA}.

Our starting point is the Hamiltonian of the electron-nuclear system
as defined in Eq.~(\ref{eq:total_hamiltonian}). This Hamiltonian is 
then split into a suitably chosen reference Hamiltonian $\op{H}_0$ and
the remainder, which is treated as a perturbation. The most appropriate reference 
system for this formalism is as follows: 
i)~The nuclear Kohn-Sham Hamiltonian~(\ref{KSNa})
rigorously defines the nuclear equilibrium positions $\underline{\bR}_0$. When expanded to 
harmonic order around these positions it can be written 
as the phonon Hamiltonian $\op{H}^{\rm ph}$ (\ref{eq:h_phonon}). 
ii)~Next we define a Born-Oppenheimer Kohn-Sham Hamiltonian via a rigid-lattice
potential given by the equilibrium coordinates $\underline{\bR}_0$,
\be
\op{H}^{\rm e}_{\rm BO} = \op{T}^{\rm e} + \op{V}^{\rm e}_{{\rm latt},\underline{\bR}_0}
+ \op{V}^{\rm e}_{\rm Hxc} + \op{\Delta}_{\rm Hxc} \,,
\ee
where
\begin{multline}
  \hat V^\text{e}_{\rm Hxc} =   \sum_{\sigma} \mint{r}
  \hat \Psi^\dagger_{\sigma}(\br)  \hat \Psi_{\sigma}(\br) \\ \times
  \left[\mint{r'} \frac{n(\br')} {|\br - \br'|} + v^\text{e}_{\rm xc}(\br)\right]
  \,,
\end{multline}
while $\op{\Delta}_{\rm Hxc}$ includes the anomalous Hartree and exchange-correlation 
contributions
\begin{multline}
  \hat \Delta_{\rm Hxc} = -\mdint{r}{r'}
  \bigg\{\hat \Psi_{\uparrow}(\br)\hat \Psi_{\downarrow}(\br') \\
  \times \left[\frac{\chi^*(\br, \br')}{|\br-\br'|} + \Delta^*_{\rm xc}(\br,\br')\right]
  + {\rm H.c.}\bigg\}
  \,.
\end{multline}

With the choice $\op{H}_0 = \op{H}^{\rm ph} + \op{H}^{\rm e}_{\rm BO}$, the interaction 
Hamiltonian reads:
\be
\label{eq:H1_def}
\op{H}_1 = \op{U}^{\rm ee} + \op{U}^{\rm e-ph}_{\rm BO} - \op{V}^{\rm n}_{\rm Hxc} - \op{V}^{\rm e}_{\rm Hxc}
-\op{\Delta}_{\rm Hxc}
\ee
The Born-Oppenheimer electron-phonon coupling operator $\op{U}^{\rm e-ph}_{\rm BO}$ 
is given by
\be
  \op{U}^{\rm e-ph}_{\rm BO} = \sum_{\sigma} \sum_{\lambda,\bq}\mint{r}
  V^{\rm BO}_{\lambda,\bq}(\br)
  \op{\Psi}^\dagger_{\sigma}(\br) \op{\Psi}_{\sigma}(\br)
  \op{\Phi}_{\lambda,\bq}
  \,.
\ee
where $V^{\rm BO}_{\lambda,\bq}(\br)$ is basically the gradient of the electronic 
Kohn-Sham potential with respect to the nuclear coordinates
and the phononic field operator is $\op{\Phi}_{\lambda,\bq} = \op{b}_{\lambda,\bq} + \op{b}_{\lambda,-\bq}^\dagger$.
Note that now we have set the auxiliary external potentials 
$\op{V}_\text{ext}^\text{e}$ and $\op{\Delta}_\text{ext}$ to zero, and
$\op{V}_\text{ext}^\text{n}$ to the bare internuclear interaction.

We believe that this is the most physical way to split the Hamiltonian,
since the electronic structure calculation for $n(\br)$, in practice, 
is usually performed for fixed nuclear positions; the nuclear
dynamics is absorbed in the exchange correlation functionals. Furthermore,
standard calculations for the electron-phonon coupling, which are based 
on linear response theory, involve the above coupling potentials 
$V^{\rm BO}_{\lambda,\bq}(\br)$.
Note that, besides the interaction terms $\op{U}^\text{ee}$ and
$\op{U}^\text{e-ph}_\text{BO}$, the perturbation includes the Hartree and
exchange-correlation potentials. In appendix~\ref{appendix-1} we
give some more details of this construction.

We now develop a many-body perturbational approach, which will
ultimately lead to explicit expressions for the exchange-correlation
functionals. The construction of this approach follows closely the
standard many-body perturbation theory described in many textbooks\cite{manybody}.
Our objective is to expand the difference $\Delta \Omega = \Omega - \Omega_\KS$ in a power
series. From this difference and with the
definitions~(\ref{eq:intomega}) and (\ref{eq:nonintomega}) it is
straightforward to derive an expression for the functional $F_{\rm xc}$.

In standard perturbation theory $\Delta \Omega$ is written as an expansion in
powers of $e^2$ and $g^2$, where $e$ (the electron charge) and $g$
(the electron-phonon coupling constant) measure respectively the
strength of the Coulomb and of the electron-phonon interactions.  In
our approach, however, every order in perturbation theory contains
{\it all} orders in $e^2$ and $g^2$. This is due to the special form
of the perturbation $\op{H}_1$ that involves the exchange-correlation
potentials which contain implicitly all orders in the interactions.
Therefore, for book-keeping purposes, we multiply the perturbation
$\op{H}_1$ by a dimensionless parameter $\Lambda$. In this way, the terms
appearing in the perturbative expansion can be labeled by powers of
$\Lambda$.

The grand canonical potential $\Omega$ can be written as
\be
  \Omega = -\frac{1}{\beta}\log(Z)
  \,,
\ee
where the partition function has the definition $Z = \Tr{\exp(-\beta \op{H})}$.
From this expression it follows directly that
\be
  \Omega-\Omega_s = -\frac{1}{\beta}\log\left(\frac{Z}{Z_\KS}\right)
  \,,
\ee
where $Z_\KS$ is the partition function of the Kohn-Sham system. It is
then straightforward, using the standard machinery of many-body
perturbation theory, to write the ratio $Z/Z_\KS$ as a series
expansion in $\Lambda$ which can be evaluated with the help of Wick's
theorem. Moreover, the number of terms in the series can be reduced by
using a generalization of the linked-cluster
theorem\cite{pGoldstone1957}. The final result reads
\be
  \Omega-\Omega_s = -\frac{1}{\beta}\sum (\text{all connected diagrams})
  \,.
\ee
In this expression the sum runs over all connected Feynman diagrams
that are topologically distinct. (Alternatively, one can expand
diagrammatically the propagators and use the Galitskii-Migdal formula
to evaluate the energy~\cite{pGalitskii1960}. The two approaches are
equivalent.)

There are several Kohn-Sham propagators that enter the Feynman
diagrams: First we have the contraction that reduces to the usual
Green's function for systems that are not superconducting 
\be
  \label{eq:KSGreenFunctG}
  G^\KS_{\sigma \sigma'}(\br\tau,\br'\tau') = -\langle{\cal \op T}
    \hat\psi_\sigma(\br\tau) \hat\psi^\dagger_{\sigma'}(\br'\tau') \rangle_\KS
  \,,
\ee
where ${\cal \op T}$ is the time-ordering operator, which orders the
operators from right to left in ascending imaginary time order, and
the average $\langle\cdots\rangle_\KS$ is done with respect to the Kohn-Sham
statistical density operator $\op{\rho}_\KS$.  This Green's function is
represented in the Feynman diagrams by a line with two arrows pointing
in the same direction.  Furthermore, due to the presence of pairing
fields in the Kohn-Sham system~(\ref{eq:nonintomega}), the
following (anomalous) averages are non-vanishing for superconducting
systems
\begin{subequations}
\label{eq:KSGreenFunctF}
\bea
  F^\KS_{\sigma \sigma'}(\br\tau,\br'\tau') & = & -\langle{\cal \op T}
    \hat\psi_\sigma(\br\tau) \hat\psi_{\sigma'}(\br'\tau')
  \rangle_\KS \,,
  \\
  F^{\KS\;\dagger}_{\sigma \sigma'}(\br\tau,\br'\tau') & = & -\langle{\cal \op T}
    \hat\psi^\dagger_\sigma(\br\tau) \hat\psi^\dagger_{\sigma'}(\br'\tau')
  \rangle_\KS
  \,.
\eea
\end{subequations}
In Feynman diagrams these propagators appear as lines with two arrows
pointing outwards (for $F^\KS$) and as lines with two arrows pointing
inwards (for $F^{\KS\;\dagger}$).  (The Green's
function~(\ref{eq:KSGreenFunctG}) and the anomalous averages
(\ref{eq:KSGreenFunctF}) can be conveniently assembled into a matrix
Green's function in Nambu-Gorkov space~\cite{pNambu1960}.)  Finally,
as the perturbation includes the electron-phonon interaction term
$\op{H}_{\rm e-ph}$, some diagrams contain the phonon propagator
(represented as a curly line)
\be
  D^\KS_{\lambda,\bq}(\tau,\tau')=\langle{\cal \op T}
    \hat \Phi_{\lambda,\bq}(\tau)\hat \Phi^\dagger_{\lambda,\bq}(\tau')
  \rangle_\KS \,.
\ee
As usual in finite temperature many-body theory, it is convenient to
work in imaginary frequency space. The time Fourier transform of the
Green's function~(\ref{eq:KSGreenFunctF}) is defined as
\be
  G^\KS_{\sigma\sigma'}(\br\tau,\br'\tau') =  
  \frac{1}{\beta}\sum_{\omega_n} \E^{-\I\omega_n(\tau-\tau')} 
  G^\KS_{\sigma\sigma'}(\br,\br';\omega_n)
  \,,
\ee
where $\omega_n = (2n+1)\pi/\beta$ are the odd Matsubara frequencies. The
frequency dependent anomalous propagators have a similar definition.
In Matsubara space the Kohn-Sham propagators read, in terms of the
Kohn-Sham particle and hole amplitudes and of the Kohn-Sham
eigenenergies,
\begin{subequations}
\begin{multline}
  G^\KS_{\sigma,\sigma'}(\br,\br';\omega_n) = \delta_{\sigma,\sigma'} \\
  \times \sum_i \left[ 
    \frac{u_i(\br) u^*_i(\br')}{\I \omega_n - E_i} +
    \frac{v_i(\br) v^*_i(\br')}{\I \omega_n + E_i} \right]
  \,,
\end{multline}
\begin{multline}
  F^\KS_{\sigma,\sigma'}(\br,\br';\omega_n) = \delta_{\sigma,-\sigma'} {\rm sgn}(\sigma') \\
  \times \sum_i \left[
    \frac{v^*_i(\br) u_i(\br')}{\I \omega_n + E_i} -
    \frac{u_i(\br) v^*_i(\br')}{\I \omega_n - E_i} \right]
  \,,
\end{multline}
\begin{multline}
  F^{\KS\; \dagger}_{\sigma,\sigma'}(\br,\br';\omega_n) = \delta_{\sigma,-\sigma'} {\rm sgn}(\sigma) \\
  \times \sum_i \left[
    \frac{u^*_i(\br) v_i(\br')}{\I \omega_n + E_i} -
    \frac{v_i(\br) u^*_i(\br')}{\I \omega_n - E_i} \right]
  \,.
\end{multline}
\end{subequations}
On the other hand, the phonon propagator depends on the even Matsubara frequencies
$\nu_n = 2n\pi/\beta$,
\be
  D^\KS_{\lambda,\bq}(\nu_n) =
  - \frac{ 2 \Omega_{\lambda,\bq}}{\nu_n^2 + \Omega_{\lambda,\bq}^2} .
\ee

\begin{figure}
\begin{tabular}{ccc}
\includegraphics[scale=1]{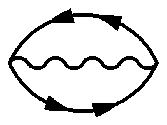} &
\includegraphics[scale=1]{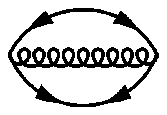} &
\includegraphics[scale=1]{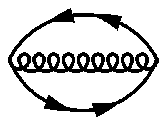} \\
a & b & c
\end{tabular}
\caption{Lowest order electronic (a) and phononic (b, c) contributions to $F_{\rm xc}$.
\label{fig:diagrams}}
\end{figure}
In first order in $\Lambda$ there is only one diagram contributing to
$F_{\rm xc}$. This diagram, depicted in Fig.~\ref{fig:diagrams}a, is
of purely electronic origin and has the same form as the standard
exchange diagram. (The wavy line in Fig.~\ref{fig:diagrams}a
represents the Coulomb interaction.) This term can be written as (for
simplicity we write all contributions to $F_{\rm xc}$ within the
decoupling approximation
\begin{multline}
  \label{eq:Fxc_a}
  F_{\rm xc}^{\rm (a)} = -\frac{1}{4}\sum_{i,j} v_{i,j}
  \left[1-\frac{\xi_{i}}{E_{i}}
  \tanh\left(\frac{\beta}{2}E_{i}\right)\right] \\
  \times \left[1-\frac{\xi_{j}}{E_{j}}
  \tanh\left(\frac{\beta}{2}E_{j}\right)\right]
  \,,
\end{multline}
where the matrix elements of the Coulomb interaction are defined as
\be
  v_{i,j} = \mdint{r}{r'}
  \varphi^*_i(\br)\varphi_i(\br')\frac{1}{|\br-\br'|}
  \varphi_j(\br)\varphi^*_j(\br')
  \,.
\ee
As the expectation value $\langle \Phi_{\lambda,\bq} \rangle = 0$, the electron-phonon
interaction does not contribute to $F_{\rm xc}$ in first order
perturbation theory in $\Lambda$. The first non-vanishing terms appear in
second order in $\Lambda$ (first order in $g^2$) and are depicted in
Fig.~\ref{fig:diagrams}b,\,c. The first of these terms
(Fig.~\ref{fig:diagrams}b) is the counterpart of the anomalous term
that contributes to the electronic Hartree
energy~(\ref{eq:el_hartree}).  Its analytic form can be written as
\begin{multline}
  \label{eq:Fxc_b}
  F_{\rm xc}^{\rm (b)} = \frac{1}{2}\sum_{\lambda,\bq} \sum_{ij}
  \left|g_{\lambda,\bq}^{ij}\right|^2 
  \frac{\Delta_i\Delta^*_j}{E_iE_j}
  \Big[I(E_i,E_j,\Omega_{\lambda,\bq}) \\ - I(E_i,-E_j,\Omega_{\lambda,\bq})\Big]
  \,,
\end{multline}
where we defined the matrix elements of the electron-phonon coupling constant
\be
  g_{\lambda,\bq}^{i j} = \mint{r} \varphi^*_i(\br) 
  V_{\lambda,\bq}^{BO}(\br)
  \varphi_j(\br)\,,
\ee
while the function $I$ is
\be
  I(E, E',\Omega) =   \frac{1}{\beta^2}\sum_{\omega_1\omega_2}
  \frac{1}{\I\omega_1 - E}\frac{1}{\I\omega_2 - E'}
  \frac{-2\Omega}{(\omega_1-\omega_2)^2  + \Omega^2}
  \,.
\ee
The three fractions in the sum come from the denominators of the two
Green's functions $G^\KS$ and from the phonon propagator $D^\KS$. It
is possible to perform the frequency sums using standard complex
contour integration methods.  The final result is
\begin{multline}
  I(E_i,E_j,\Omega) =
  f_\beta(E_i) \, f_\beta(E_j) \, n_\beta(\Omega)  \\
  \times
  \left[ \frac{\E^{\beta E_i} - \E^{\beta (E_j + \Omega)}}{E_i - E_j - \Omega}
  -  \frac{\E^{\beta E_j} - \E^{\beta (E_i + \Omega)}}{E_i - E_j + \Omega} 
  \right]
  \,.
\end{multline}
A diagram analogous to the one depicted in Fig.~\ref{fig:diagrams}b
but with the phonon propagator replaced by the bare Coulomb
interaction exists as well.  This diagram is the anomalous Hartree
term which appears as the second term on the right hand side of
Eq.~(\ref{eq:el_hartree}). Since this term is treated explicitly in the
electronic Kohn-Sham equations it is not part of the exchange-correlation functional.

Note that expression~(\ref{eq:Fxc_b}) is so much more complicated than
the anomalous Hartree term simply because the phonon propagator
describes a {\it retarded} interaction. If we assumed a retarded
electronic interaction instead of the instantaneous Coulomb potential
$1/|\br-\br'|$ the anomalous Hartree contribution would look very
similar to~(\ref{eq:Fxc_b}).

The second phononic term that contributes to $F_{\rm xc}$ is depicted
in Fig.~\ref{fig:diagrams}c. This Feynman diagram has the same
structure as the electronic exchange term (Fig.~\ref{fig:diagrams}a).
However, and again due to retardation effects, its analytic structure
is more complicated than~(\ref{eq:Fxc_a}), namely
\begin{multline}
  \label{eq:Fxc_c}
  F_{\rm xc}^{\rm (c)} = - \frac{1}{2}\sum_{\lambda,\bq} \sum_{ij}
  \left|g_{\lambda,\bq}^{ij}\right|^2 \Bigg[
    \left(1+\frac{\xi_i}{E_i}\frac{\xi_j}{E_j}\right)
    I(E_i,E_j,\Omega_{\lambda,\bq}) \\
    + \left(1-\frac{\xi_i}{E_i}\frac{\xi_j}{E_j}\right)
    I(E_i,-E_j,\Omega_{\lambda,\bq})
  \Bigg]
  \,.
\end{multline}
The self-energy diagrams contributing to $F_{\rm xc}^{\rm (b)}$ and $F_{\rm xc}^{\rm (c)}$ 
resemble the self-energy diagrams that appear in Eliashberg 
theory~\cite{Eliashberg,scalapino}. 
By virtue of Migdal's theorem\cite{migdal}, vertex corrections should be small and are
therefore neglected, both in Eliashberg theory and in our treatment. There is, however, an important
difference: in Eliashberg theory the propagators that enter the
self-energy diagrams are {\it dressed} propagators, while in our case
we have {\it bare} (Kohn-Sham) propagators. By using the bare propagators we neglect 
more diagrams than those containing vertex corrections. We postpone a more detailed discussion 
of this problem to the section on the exchange-correlation kernels.

\section{Functional derivatives and the chain rule}
\label{sec:5}

We have seen in Eq.~(\ref{eq:xc_potentials_anomalous}) how the
anomalous exchange-correlation potential is defined as the functional
derivative of the exchange-correlation free-energy functional with
respect to the anomalous density $\chi$. However, the contributions to
the exchange-correlation free-energy functional that stem from
Kohn-Sham perturbation theory are only known as explicit functionals
of the Kohn-Sham orbitals $\varphi_i(\br)$, the Kohn-Sham single-particle
energies $(\epsilon_i-\mu)$, and the pair potential $\Delta_i$. Of course, the
Hohenberg-Kohn theorem tells us that these quantities are themselves
functionals of the densities, so the free-energy is an {\it implicit}
functional of the densities. Furthermore, if one makes the additional
approximation that $v^\text{e}_{\rm xc}$ does not depend on $\chi$ then the
Kohn-Sham orbitals $\varphi_i(\br)$ and the eigenenergies $\epsilon_i$ are also
independent of the anomalous density. $F_{\rm xc}$ is then a function
of the chemical potential $\mu$ and a functional of the (complex) pair
potential $\Delta_i$
\be
  F_{\rm xc} = F_{\rm xc}[\mu, |\Delta_i|^2, \phi_i]
  \,.
\ee
For convenience, instead of working with $\Delta_i$ and $\Delta_i^*$, we prefer using the 
modulus square of the pair potential and its phase as independent variables. The 
anomalous exchange-correlation potential can thus be evaluated using the
chain rule for functional derivatives
\be
  \Delta_{{\rm xc}\;i}  =  
  - \frac{\delta F_{\rm xc}}{\delta \mu}\frac{\delta \mu}{\delta \chi^*_i}
  - \sum_{j}\left[
    \frac{\delta F_{\rm xc}}{\delta |\Delta_j|^2}\frac{\delta |\Delta_j|^2}{\delta \chi^*_i}
   +\frac{\delta F_{\rm xc}}{\delta (\phi_j)} \frac{\delta (\phi_j)}{\delta \chi^*_i}
  \right]
  \,.
\ee
The partial derivatives of $F_{\rm xc}$ can now be calculated
directly. The remaining functional derivatives are somehow harder to
obtain, but can be derived from the definitions of the densities,
equations~(\ref{eq:el-densities}), and from the fact that the particle
density and the anomalous density are independent variables. This
latter condition can be expressed through the relation
\be
  \frac{\delta n(\bx)}{\delta \chi^*(\br,\br')} = 0
  \,.
\ee
Moreover, we make use of the two trivial conditions
\be
  \frac{\delta \chi^*_i}{\delta \chi^*_j} = \delta_{i,j}
  \quad,\quad
  \frac{\delta \chi_i}{\delta \chi^*_j} = 0
  \,.
\ee
From the above conditions, and after some algebra, we arrive at the expressions for
the remaining functional derivatives
\begin{subequations}
\bea
  \frac{\delta \left|\Delta_j\right|^2}{\delta \chi^*_i}
  & = &
  \frac{2}{Y^0_j} \left[E_j^2\Delta_j \delta_{i,j} -
    Y^1_j \left|\Delta_j\right|^2 
    \frac{\delta \mu}{\delta \chi^*_i} \right] \,,
  \\
  \frac{\delta (\phi_j)}{\delta \chi^*_i} \,,
  & = &
  \I \delta_{i,j} \frac{E_i}{\Delta_i^* 
    \tanh\left(\frac{\beta}{2}E_i\right)}
  \\
  \frac{\delta \mu}{\delta \chi^*_i}
  & = &
   - {Z^1_i} \Big/ \; {\sum_j Z^0_j}
  \,.
\eea
\end{subequations}
The functions $Z^0_i$ and $Z^1_i$ have the definitions
\be
  \label{Z0}
  Z^0_i = \frac{E_i}{Y^0_i}
  \frac{\frac{\beta}{2}\tanh\left(\frac{\beta}{2} E_i\right)}
    {\cosh^2\left(\frac{\beta}{2} E_i\right)}
  \quad ; \quad
  Z^1_i = \frac{Y^1_i}{Y^0_i} \Delta_i
  \,,
\ee
and, finally, the two functions $Y^0_i$ and $Y^1_i$ read
\begin{subequations}
\bea
  Y^0_i & = &
  \frac{\xi_i^2}{E_i}\tanh\left(\frac{\beta}{2} E_i\right)
  + \frac{\frac{\beta}{2} \left|\Delta_i\right|^2}
  {\cosh^2\left(\frac{\beta}{2} E_i\right)} \,,
  \\
  Y^1_i & = &
  \frac{\xi_i}{E_i}\tanh\left(\frac{\beta}{2} E_i\right)
  - \frac{\frac{\beta}{2} \xi_i}
  {\cosh^2\left(\frac{\beta}{2} E_i\right)}
  \,.
\eea
\end{subequations}

There exists another method to obtain exchange-correlation functionals
using Kohn-Sham perturbation theory {\it without} resorting to functional 
derivatives. This method follows the ideas of Sham and Schl{\"u}ter~\cite{SHSCL}, and 
provides a very direct connection between many-body perturbation theory and the
normal and anomalous exchange-correlation functionals. In the following we will
give a brief account of the Sham-Schl{\"u}ter method for superconductors.

There is a simple connection between the electron density $n(\br)$ and the interacting
many-body Green's function
\be
  \label{FullDens1}
  n(\br) = \lim_{\eta\to 0^+} 
  \frac{1}{\beta}\sum_{\omega_n}\sum_\sigma
  \E^{\I\eta\omega_n}G_{\sigma\sigma}(\br,\br;\omega_n)
  \,.
\ee
The definition of the interacting Green's function is similar to
Eq.~(\ref{eq:KSGreenFunctG}), but with the thermal average weighted by
the interacting statistical operator $\op{\rho}_0$.  The infinitesimal
$\eta$ is used to ensure the correct ordering of the field operators in
Eq.~(\ref{eq:KSGreenFunctG}) so that Eq.~(\ref{FullDens1}) is
fulfilled. In the same way it is simple to prove that the anomalous
density obeys the relation
\be
  \label{FullDens2}
  \chi(\br,\br') = - \lim_{\eta\to 0^+} 
  \frac{1}{\beta}\sum_{\omega_n}
  \E^{\I\eta\omega_n}F_{\uparrow\downarrow}(\br,\br';-\omega_n)
  \,,
\ee
where $F$ is the interacting anomalous propagator. On the other hand, we defined 
the Kohn-Sham system as the \emph{non-interacting} system whose both normal
and anomalous densities are equal to the densities of the \emph{interacting} 
system. Therefore, one can equally calculate the interacting densities
from the Kohn-Sham Green's functions
\bea
  \label{KSDens1}
  n(\br) = \lim_{\eta\to 0^+} 
  \frac{1}{\beta}\sum_{\omega_n}\sum_\sigma
  \E^{\I\eta\omega_n}G^\KS_{\sigma\sigma}(\br,\br;\omega_n)
  \,,
\eea
with an equation similar to Eq.~(\ref{FullDens2}) relating $\chi$ and
$F^\KS$.  We then expand perturbatively the interacting Green's
functions in~(\ref{FullDens1}) and ~(\ref{FullDens2}), and equate the
two equations for $n(\br)$, and the two equations for $\chi(\br,\br')$.
As the perturbation~(\ref{eq:H1_def}) includes the terms
$\op{V}^\text{e}_{\rm xc}$ and $\op{\Delta}_{\rm xc}$, the resulting
equations form a system of two coupled integral equations that allow
the determination of $v^\text{e}_{\rm xc}$ and $\Delta_{\rm xc}$.  Those
integral equations are the so-called Sham-Schl{\"u}ter equations.

The two methods to obtain the exchange-correlation potentials are
conceptually quite different. The first uses the
definitions~(\ref{eq:xc_potentials}) to derive the
exchange-correlation potentials using a series of chain rules. The
Sham-Schl{\"u}ter approach is closer to many-body perturbation theory, and
provides a natural relationship between the Green's function and the
exchange-correlation potentials of DFT. However, both approaches lead
to the same result if the free energy in the first method is expanded
up to the same order in perturbation theory as the Green's functions
in the second method.

\section{A simple BCS-like model}
\label{sec:6}

We now introduce a simple model that will allow us to study in detail
the functionals developed in this article. For simplicity, we assume
that the pair potential has $s$-wave symmetry and behaves
approximately like $\Delta_i = \Delta(\xi_i)$.  This assumption is fulfilled by
all traditional superconductors.  In this model, we can transform the
gap equation into a one-dimensional equation in energy space
\begin{multline}
  \label{eq:gap_E}
  \Delta(\xi) = -{\cal Z}(\xi) \Delta(\xi)\\ - \frac{1}{2}\int_{-\mu}^{\infty} d\xi' N(\xi') 
  {\cal K}(\xi,\xi') \frac{\tanh\left(\frac{\beta}{2}\xi'\right)}{\xi'} \Delta(\xi')\,, 
\end{multline}
where $N(\xi)$ is the density of states at the energy $\xi$.  It is
possible to further simplify this equation by assuming a BCS-like
model for both ${\cal K}$ and ${\cal Z}$.  If we assume that the
kernel ${\cal K}$ and the renormalization term ${\cal Z}$ are constant 
in a shell of width $\omega_c$ around the Fermi energy and zero outside 
this region, Eq.~(\ref{eq:gap_E}) can be solved analytically for the
transition temperature $T_{\rm c}$
\be
  \label{eq:tc_model}
  T_{\rm c} \propto
  \exp\left[\frac{1+{\cal Z}(0)}{N(0){\cal K}(0)}\right]
  \,,
\ee
where ${\cal K}(0)$ and ${\cal Z}(0)$ are the values of ${\cal
  K}(\xi,\xi')$ and ${\cal Z}(\xi)$ at the Fermi surface.
Equation~(\ref{eq:tc_model}) has exactly the same structure as
McMillan's formula\cite{mcmillan,pCarbote1990}, which is an approximate solution of the Eliashberg
equations. This latter formula reads
\be
  \label{McMillan}
  T_c = \frac{\Omega_\text{log}}{1.20} \exp\left[-\frac{1.04(1+\lambda)}{\lambda-\mu^*(1+0.62\lambda)}\right]
  \,.
\ee
The number $\mu^*$, the Coulomb pseudo-potential of Eliashberg theory, measures 
the strength of the electron-electron interaction. This parameter is quite hard to 
calculate and is often fitted to experimental data.  As $\mu^*$ is normally
positive, it tends to decrease the superconducting transition temperature.
On the other hand, $\lambda$ is a measure of the electron-phonon coupling strength
\be
  \lambda = 2 \int \D\Omega\; \frac{\alpha^2 F(\Omega)}{\Omega}
  \,.
\ee
The behavior of $T_{\rm c}$ with $\lambda$ is very non-linear: For small values of
$\lambda$, $T_{\rm c}$ grows exponentially; However, as $\lambda$
increases, the superconducting transition temperature saturates. 
The parameter $\Omega_\text{log}$ is a weighted average of the phonon frequencies
\be
  \Omega_\text{log} = \exp\left[\frac{2}{\lambda}\int \D\Omega\; \log(\Omega) \frac{\alpha^2 F(\Omega)}{\Omega}
        \right]
  \,,
\ee
and is of the order of the Debye frequency of the material.
Finally, the Eliashberg spectral function is the electron-phonon coupling
constant averaged on the Fermi surface
\be
  \alpha^2F(\Omega) = \frac{1}{N(0)} \sum_{ij}\sum_{\lambda,\bq} 
  \left| g_{\lambda,\bq}^{ij} \right|
  \delta(\xi_i) \delta(\xi_j)\delta(\Omega - \Omega_{\lambda,\bq})
\ee
It is widely accepted that McMillan's formula gives a quite
accurate description of the transition temperature for simple, BCS-like,
superconductors. Therefore, by comparing expressions~(\ref{eq:tc_model}) and
(\ref{McMillan}) for the phonon-only case, i.e. $\mu*=0$, we obtain that for 
BCS-like superconductors
\be
  \label{eq:mc_scale}
  N(0){\cal K}(0) \approx -\lambda  \quad;\quad {\cal Z}(0) \approx \lambda
  \,.
\ee
This is an extremely important property of the exchange-correlation kernel, 
which should be fulfilled by any approximate functional.

\section{Approximations to the anomalous Hartree exchange correlation kernel}
\label{sec:7}

From the perturbative expansion of the exchange-correlation
free-energy it is clear that we can split the free-energy in three
parts: The first contains the purely electronic terms, i.e., the terms
that do not contain explicitly the electron-phonon coupling constant;
The second, terms only involving the electron-phonon coupling
constant; And the last, which we define as the total free-energy minus
the two first parts, will have mixed contributions from the Coulomb
and electron-phonon interactions. The exchange-correlation potentials
and the exchange-correlation kernels can be split in the same way.

In this section we develop exchange-correlation kernels to be used in
the linearized gap equation~(\ref{eq:gap}). Functionals that can be
used in the non-linear gap equation~(\ref{eq:nlgap}) are discussed
later. This section is divided in two parts: First we look at the
purely electron-phonon contributions to the exchange-correlation
kernel. Such functionals are developed using the machinery of
Kohn-Sham perturbation theory together with the chain rule introduced
earlier. In the second part, we turn our attention to the purely
electronic part of the kernel.  Two functionals will be presented: the
first has the form of an local density approximation (LDA), 
while the second is a functional that
avoids the direct computation of the screened Coulomb matrix elements.
The mixed contributions appearing in the perturbational expansion of
the free energy are neglected in the current treatment.

\subsection{Electron-phonon contributions}

In first order in $g^2$ there are two terms stemming from the electron-phonon interaction
that contribute to the exchange-correlation free energy: $F_{\rm xc}^{\rm (b)}$ given by
equation~(\ref{eq:Fxc_b}), and  $F_{\rm xc}^{\rm (c)}$ given by
equation~(\ref{eq:Fxc_c}). The exchange-correlation kernel derived from 
$F_{\rm xc}^{\rm (b)}$ is non-diagonal and has the form 
\begin{multline}
  \label{eq:kph_ij}
  {\cal K}^{\rm ph}_{ij} = \frac{2}{\tanh\bek{\xi_i}\tanh\bek{\xi_j}}
  \sum_{\lambda,\bq} \left|g_{\lambda,\bq}^{ij}\right|^2 
   \\ \times \left[
    I(\xi_i,\xi_j,\Omega_{\lambda,\bq}) - I(\xi_i,-\xi_j,\Omega_{\lambda,\bq})
  \right] \,.
\end{multline}

\begin{figure}
  \centerline{%
    \includegraphics[scale=1]{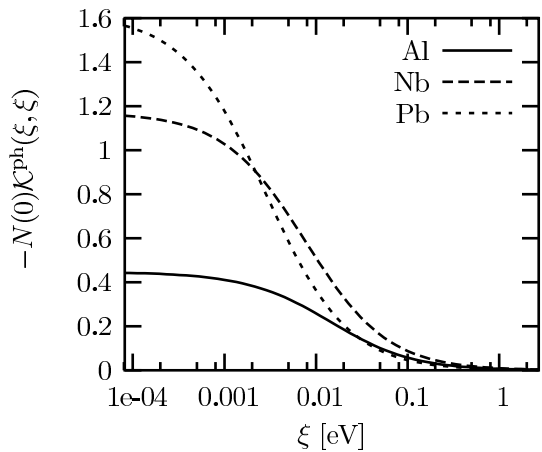}%
  }
  \caption{The function $-N(0){\cal K}^{\rm ph}(\xi,\xi)$ for Al, Nb and Pb, calculated at $T=0$\,K
  \label{fig:kphonon}
  }
\end{figure}
To gain further insight into this term, we use a simplified model: we
approximate the electron-phonon coupling constants by their average
value at the Fermi surface and the electronic energy dispersion is
replaced by the free-electron model.  In Fig.~\ref{fig:kphonon} we
depict the diagonal ${\cal K}^{\rm ph}(\xi,\xi)$ for aluminum, niobium
and lead at zero temperature for this simplified model.  As this 
contribution to the 
exchange-correlation kernel exhibits particle-hole symmetry we
only plot the region $\xi > 0$. This term is sharply peaked at the Fermi
energy (note the logarithmic scale in the $\xi$-axis). Furthermore, the
width of the curves for each material is of the order of the
corresponding Debye frequency. The value of the kernel at the Fermi
energy can be calculated analytically
\begin{multline}
  N(0){\cal K}^{\rm ph}(0,0) = 
    - \int\! \D\Omega\: \alpha^2F(\Omega) \\
    \times \frac{2}{\Omega} 
    \left[ 1 - \frac{4}{\beta\Omega}
    {\rm cotanh}\left(\frac{\beta\Omega}{2}\right)
    + \frac{8}{\left(\beta\Omega\right)^2}\right] \,.
\end{multline}
At zero temperature, the value of $N(0){\cal K}^{\rm ph}(0,0)$ reduces
to $-\lambda$, which is the value expected from the comparison to McMillan's
formula [cf. Eq.~(\ref{eq:mc_scale})].  However, at higher temperature
$N(0){\cal K}^{\rm ph}(0,0)$ decreases monotonically.

The second phononic contribution to the exchange-correlation kernel
coming from the Kohn-Sham perturbation theory (PT) originates from 
the diagram $F_{\rm xc}^{\rm (c)}$. It is a diagonal term, that reads
\begin{widetext}
\begin{multline}
  \label{eq:Z0}
  {\cal Z}^{\rm ph,PT}_{i} =  - \frac{2}{\sum_j
  \frac{\beta/2}{\cosh^2\left(\frac{\beta}{2} \xi_j\right)}}
  \left[\frac{1}{\xi_i} -\frac{\beta / 2}{
  \sinh\left(\frac{\beta}{2} \xi_i\right)
  \cosh\left(\frac{\beta}{2} \xi_i\right)}\right]
  \sum_{jl} \sum_{\lambda,\bq} \left| g^{jl}_{\lambda,\bq}\right|^2
  I'(\xi_j,\xi_l,\Omega_{\lambda,\bq})
  \\  + 
  \frac{1}{\tanh\left(\frac{\beta}{2} \xi_i\right)}
  \sum_{j} \sum_{\lambda,\bq} \left| g^{ij}_{\lambda,\bq}\right|^2
  \Bigg\{ \frac{1}{\xi_i}
  \left[I(\xi_i,\xi_j,\Omega_{\lambda,\bq}) - 
    I(\xi_i,-\xi_j,\Omega_{\lambda,\bq})\right]
  - 2 I'(\xi_i,\xi_j,\Omega_{\lambda,\bq})\Bigg\}
  \,,
\end{multline}
\end{widetext}
where the function $I'$ is defined as 
\be I'(\xi_i,\xi_j,\Omega_{\lambda,\bq}) =
   \frac{\partial}{\partial \xi_i} I(\xi_i,\xi_j,\Omega_{\lambda,\bq}) \,.  
\ee 
If we try to apply the simplified model presented earlier we find that
${\cal Z}^{\rm ph,PT}_{i}$ diverges logarithmically. This divergence can
be traced back to the substitution of $g^{ij}_{\lambda,\bq}$ by its value at the
Fermi surface. This problem can be solved by retaining the full
dependence of the electron-phonon coupling constant on the indexes
$i$, and $j$: $g^{ij}_{\lambda,\bq}$ then decays as a function of energy
thereby making the integrals present in Eq.~(\ref{eq:Z0}) convergent.
A closer analysis of the expressions also reveals that the divergent
part of the integrands is antisymmetric around the Fermi surface.
Therefore, the divergent integrals would vanish in the case of
particle-hole symmetry.  It seems then reasonable to neglect the
antisymmetric part of the integrands, retaining only the symmetric
part. The new functional reads
\begin{multline}
  \label{Dterm147}
  {\cal Z}^\text{ph, sym}_i = 
  \frac{1}{\tanh\left(\frac{\beta}{2} \xi_i\right)}
  \sum_{j} \sum_{\lambda,\bq} \left| g^{i,j}_{\lambda,\bq}\right|^2
  \\ \times
  \left[I'(\xi_i,\xi_j,\Omega)+I'(\xi_i,-\xi_j,\Omega)\right]
  \,.
\end{multline}
\begin{figure}
  \centerline{%
    \includegraphics[scale=1]{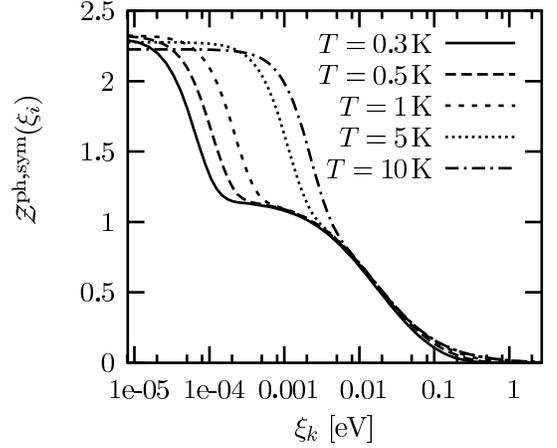}%
  }
  \caption{The dependence of ${\cal Z}^\text{ph, sym}_{i}$ on temperature for niobium.
  \label{fig:zphonon}
  }
\end{figure}
\begin{table}
\begin{center}
\begin{tabular}{c|cccccc}
    &  Al  &  Nb  &  Mo  &  Ta  &  V   &  Pb   \\ \hline
${\cal K}^{\rm ph} + {\cal Z}^\text{ph, sym}$
    & 5.59 & 15.7 & 4.14 & 8.48 & 23.2 & 8.12 \\ 
${\cal K}^{\rm ph} + {\cal Z}^\text{ph}$
    & 7.10 & 23.0 & 5.23 & 11.7 & 34.2 & 12.8 \\
Eliashberg   
    & 9.75 & 24.7 & 7.31 & 14.0 & 36.4 & 12.2 \\ 
\end{tabular}
\end{center}
\caption{Transition temperatures from numerical solutions of the phonon-only DFT and 
  Eliashberg equations. All temperatures are in Kelvin.
\label{phelTcs}}
\end{table}
In Fig.~\ref{fig:zphonon} this term is plotted for niobium for several
temperatures. It turns out that the function ${\cal Z}^\text{ph,
  sym}(\xi_i)$ is a smooth function of the energy, and its value at the
Fermi surface ($\xi_k=0$) is approximately $2 \lambda$. This is twice the
value expected from the comparison to McMillan's formula [cf.
Eq.~(\ref{eq:mc_scale})].  Furthermore, a careful analysis of
Fig.~\ref{fig:zphonon} suggest that ${\cal Z}^\text{ph, sym}(\xi_i)$ can
be written as the sum of two terms: i)~one broader and very weekly
temperature dependent; ii)~a second contribution whose width decreases
significantly with the temperature. Both terms contribute with
approximately $\lambda$ to the value of ${\cal Z}^\text{ph, sym}(\xi_i)$ at
the Fermi surface. As the renormalization term ${\cal Z}^\text{ph,
  sym}(\xi_i)$ appears to be too large, one can expect that transition
temperatures calculated with this functional will be too small. The
situation should be worst for the strong-coupling superconductors like
niobium or lead, where the renormalization is large. This is confirmed
by Table~\ref{phelTcs} where we list the transition temperatures
obtained with the phononic part of the functional.  These numbers are
compared to solutions of Eliashberg's equation where we neglected the
electron-electron repulsion.

We believe that the shortcomings of this functional can be traced back
to the following: Migdal's theorem tells us that, to a very good
approximation, we can neglect in the perturbative expansion diagrams
including vertex corrections due to the electron-phonon interaction.
However, diagrams including self-energy insertions of phononic origin
should be included to have a consistent description of the electron-phonon interaction. 
Therefore the {\it bare} Green's functions entering in the diagrams depicted in
Fig.~\ref{fig:diagrams}b,c should be replaced by {\it dressed}
propagators. In a first step to improve our functionals we dressed the
propagators with a subset of phonon self-energy insertions. We found
that the non-diagonal term ${\cal K}^{\rm ph}$ is basically insensitive,
while the term ${\cal Z}^\text{ph, sym}$ is reduced by roughly 20\%. This is
almost half the correction necessary to satisfy Eq.~(\ref{eq:mc_scale}).
We expect that the other 30\% are accounted for by the remaining 
self-energy insertions.  However, this approach is quite involved
numerically, so we choose a different path to improve our functional.

We know that the phonon renormalization term should have the value $\lambda$
at the Fermi surface. Furthermore, this term should have a width
comparable to the Debye frequency.  It is clear that the broader
contribution to ${\cal Z}^\text{ph, sym}(\xi_i)$ obeys these
requirements. We therefore propose to separate the two parts of
${\cal Z}^\text{ph, sym}(\xi_i)$ and use the part i) as our renormalization
term. We believe that this procedure is at least partially justified
by the results obtained by dressing the Green's functions. The
functional corrected in this way reads
\begin{multline}
\label{eq:Z_ph}
  {\cal Z}^\text{ph}_i = \frac{1}{\tanh\left(\frac{\beta}{2} \xi_i\right)}
  \sum_{j} \sum_{\lambda,\bq} \left| g^{ij}_{\lambda,\bq}\right|^2 \\
  \left[J(\xi_i,\xi_j,\Omega_{\lambda,\bq})+J(\xi_i,-\xi_j,\Omega_{\lambda,\bq})\right]
  \,,
\end{multline}
where the function $J$ is defined by
\be
  J(\xi,\xi',\Omega) = \tilde J(\xi,\xi',\Omega) - \tilde J(\xi,\xi',-\Omega)
  \,.
\ee
Finally we have
\begin{multline}
  \tilde J(\xi,\xi',\Omega) = - \frac{f_\beta(\xi) + n_\beta(\Omega)}{\xi-\xi'-\Omega} \\
  \times \left[
  \frac{f_\beta(\xi')-f_\beta(\xi-\Omega)}{\xi-\xi'-\Omega}
  -\beta f_\beta(\xi-\Omega)f_\beta(-\xi'+\Omega)
  \right]\,.
\end{multline}
The functional ${\cal Z}^\text{ph}$ is smooth both as a function of the
energy and as a function of the temperature. Furthermore, it has
approximately the value $\lambda$ at the Fermi surface.  
The functional (\ref{eq:Z_ph}), together with the phononic 
kernel (\ref{eq:kph_ij}), is a central result of our work.
It is the functional that will be used in the calculations of II.
In Table~\ref{phelTcs}, we present the phonon-only transition temperatures
calculated with this functional. All $T_\text{c}$'s are in quite
good agreement with transition temperatures calculated from
Eliashberg's equation. We emphasize that the transition temperatures
in Table~\ref{phelTcs} are given exclusively for the purpose of testing
and/or calibrating the approximations made for the phononic part of
the exchange-correlation functional. $T_\text{c}$'s resulting from 
setting $\mu^*=0$ in the Eliashberg
equations and setting the Coulomb terms to zero in the DFT context
have, of course, nothing to do with the $T_c$'s observed in nature.
For results including the Coulomb terms,  we refer the reader to II.

\subsection{Electron-electron contributions}

We now develop functionals that take into account the Coulombic part
of the interaction.  There are two terms in the energy functional that
give contributions to the linearized gap-equation~(\ref{eq:gap}). The first is
the anomalous contribution to the Hartree energy, given
Eq.~(\ref{eq:el_hartree}), and the second is the exchange term
$F_\text{xc}^\text{a}$ depicted in Fig.~\ref{fig:diagrams}a. The
interaction that enters these expressions is the bare Coulomb
interaction, $1/|\br-\br'|$. However, electrons in a metal do not
feel the bare Coulomb interaction, but a much weaker interaction,
screened by the sea of electrons.  To take this into account, we take
a step back, and propose an alternative form to the energy functional
based on the superconducting version of the local density
approximation (LDA)~\cite{SC-LDA}. In this approach the
exchange-correlation energy of the inhomogeneous system is written in
terms of the exchange-correlation energy density of the homogeneous
superconducting electron-gas
\begin{multline}
  \label{SCLDA}
  F_{\rm xc}^{\rm SCLDA}[n(\bR), \chi(\bR,\bk)] = \\
  \mint{R}
  \left.f_{\rm xc}^{\rm hom}[n, \chi(\bk)]\right|_{\substack{
  n = n(\bR) \\ \chi = \chi_W(\bR,\bk)}}
  \,,
\end{multline}
where $\chi_W(\bR,\bk)$ is the Wigner transform of the anomalous density
of the inhomogeneous system, given by
\be
  \chi_W(\bR,\bk) = \mint{s} \E^{\I\bk\bs}
  \chi\left(\bR+\frac{\bs}{2}, \bR-\frac{\bs}{2}\right)
  \,.
\ee
It is easy to see that this definition reduces to the usual LDA for
non-superconducting systems in the limit $\chi \to 0$. Moreover, it is
possible to prove that this is the only consistent definition of an
LDA for the superconducting state~\cite{pUllrich1995}.  As an
approximation to the exchange-correlation energy of the electron-gas
one could take the random phase approximation (RPA) functional
proposed in Ref.~\onlinecite{SC-LDA}. However, this functional has
only been evaluated for a very simple class of pair potentials, namely
Gaussians centered at the Fermi surface. We therefore propose an
alternative and simpler form to the Coulombic contribution to $F_{\rm xc}$. 
For convenience, we approximate {\it together} the anomalous
Hartree and the exchange-correlation contributions.  Our approximation
reads
\bea
  \label{fxchom_TF}
  \lefteqn{f_\text{Hxc}^\text{hom}[\chi](n) - f_\text{xc}^{\text{hom,NS}}(n) =} \\&&
  \mint{(r-r')} |\chi(\br-\br')|^2 v^{\rm TF}(\br-\br') \nonumber
  \,,
\eea
where $v^{\rm TF}(\br-\br')$ is the Coulomb interaction screened by a 
Thomas-Fermi model. In coordinate space the Thomas-Fermi interaction reads
\be
  v^{\rm TF}(\br-\br') = \frac{e^{-k_{\text{TF}} |\br-\br'| }}{|\br - \br'|}
  \,,
\ee
with the Thomas-Fermi screening length, $k_{\rm TF}$, given by
\be
  k^2_{\rm TF} = 4 \pi N(0) \,.
\ee
By inserting expression~(\ref{fxchom_TF}) in the definition of the LDA, 
equation~(\ref{SCLDA}), we can identify this approximation as a 
{\it Thomas-Fermi screened anomalous Hartree} contribution to the free-energy.

The anomalous Hartree exchange-correlation kernel stemming from this term is simply
\be
  {\cal K}^\text{TF}_{ij} = v^\text{TF}_{ij}
  \,,
\ee
where the matrix elements of the Thomas-Fermi interaction are defined by
\be
  \label{eq:vTF_ij}
  v^\text{TF}_{ij} = \mdint{r}{r'} \varphi^*_i(\br)\varphi_i(\br') v^{\rm TF}(\br-\br')
  \varphi_j(\br)\varphi^*_j(\br')
  \,
\ee
where the $\varphi_i$'s are the Kohn-Sham orbitals of the inhomogeneous system at hand.

In II we will compare the results obtained with the above approximation
with further simplified expressions.
In the simplest model, the Kohn-Sham orbitals are taken to be plane waves 
with a parabolic dispersion. In this case, the kernel can be written in 
energy space (after averaging over the angles) as:
\be
  \label{eq:vTF_e}
  {\cal K}_{\rm TF}(\xi,\xi') = \frac{\pi}{k k'}
  \log \left[\frac{\left(k+k'\right)^2+k^2_{\rm TF}}
    {\left(k-k'\right)^2+k^2_{\rm TF}}\right]
  \,,
\ee
with $k=\sqrt{2(\xi-\mu)}$ and $k'=\sqrt{2(\xi'-\mu)}$. Using the BCS-like two-well model
one can extract the counterpart of the Coulomb pseudo-potential $\mu^*$ from Eliashberg
theory. A crude estimate for $r_s=2$ gives a value around 0.1, which compares well with
the typical values of $\mu^*$ for simple metals ($\mu^*=$~0.10 -- 0.15)~\cite{pCarbote1990}. 
It should be stressed again at this point, that the present method does not
require $\mu^*$. The estimates given here are used to demonstrate to which
values of $\mu^*$ our ab initio Coulomb terms correspond to. 

While the replacement of the Kohn-Sham orbitals in (\ref{eq:vTF_ij})
by plane waves may be acceptable for simple metals, it will be too
crude for more complex materials. In those cases it is still possible
to avoid the direct computation of the screened Coulomb matrix
elements~(\ref{eq:vTF_ij}), by going along the lines described by Sham
and Kohn~\cite{pSham1966}.  We briefly outline here the main points of
this classical paper, which deals with an approximate way of getting
an electron self-energy for the normal state. We assume, as usual
within the LDA, that our system can be described around the point
$\br$ by a homogeneous electron gas of density $n(\br)$. The
wave-functions of this electron gas can be locally expressed as
plane-waves of momentum $\bp(\br)$ whose value is determined, in a
semi-classical way, from the electron energy of the real system. In
the simplest form, the mapping can be obtained from Eqs.~(4.5) and (4.13)
of Ref.~\onlinecite{pSham1966} as
\be
  \label{eq:map}
  \frac{p^2}{2} = \xi_i + \mu_h(n(\br)) \,, 
\ee 
where $\mu_h(n)$ is the chemical potential of an non-interacting
homogeneous electron gas with density $n$.  Furthermore, we
approximate $\mu_h(n(\br))$ by the constant $\mu_h(n)$, where $n$ is the
average density of the material.  We suggest here to approximate the
Coulomb interaction kernel between electrons at energies $\varepsilon_{\bk}$ and
$\varepsilon_{\bk'}$ by the corresponding quantities in the free-electron gas.
We then replace $p^2/2 \to \xi_i+\mu_h = \eta_i$, and rewrite the
interaction~(\ref{eq:vTF_e}) as
\be 
{\cal K}^\text{SK}_{i,j} = \frac{\pi}{2 \sqrt{\eta_i
    \eta_j}}\log\left(\frac{\eta_i+\eta_j+2\sqrt{\eta_i~\eta_j}+k^2_\text{TF}/2}
  {\eta_i+\eta_j-2\sqrt{\eta_i \eta_j}+k^2_\text{TF}/2}\right) \,.  
\ee 
In principle, one could consider not only $p$ but also $k_\text{TF}$
as locally dependent on the density $n(\br)$. In our simplified
approach, however, we fix the Thomas-Fermi screening length to a
constant value.  

Eq.~(\ref{eq:map}) is conceived in terms of wave packets, and is valid
if $n(\br)$ does not vary too much on the scale of the Fermi length,
exactly as in the normal state LDA.  One can speculate, however, that
when applied to the superconducting state the relevant length scale
becomes the coherence length, normally much larger than the atomic
scale. Therefore, we may assume that local variations of the density
on the atomic scale will not affect the final superconducting
properties.

It should be noted that this approximation, although derived in the 
spirit of an LDA, is not a local density approximation, since it does
not depend explicitly on the densities, but {\it implicitly} via the
single-particle energies $\xi_i$.

\section{Functionals for the non-linear gap equation}
\label{sec:8}

In this section we provide approximations to the exchange-correlation
kernel that can be used in the non-linear
gap-equation~(\ref{eq:nlgap}). These functionals will obey to one
constraint, namely that upon linearization they will reduce to the
functionals presented in the previous section. This assures that the
gap functions obtained from Eq.~(\ref{eq:nlgap}) and the transition
temperatures calculated from Eq.~(\ref{eq:nlgap}) are consistent.
Furthermore, we require these functionals to be ``well behaved'',
i.e., without discontinuities or any other kind of pathological
behaviors.

The simplest way to derive an exchange-correlation functional is to
use the expressions derived through Kohn-Sham perturbation theory in
the definition~(\ref{eq:xc_potentials_anomalous}).  For example, the
first phononic contribution $F_{\rm xc}^{\rm (b)}$
(Fig.~\ref{fig:diagrams}b) yields the contribution
\begin{widetext}
\begin{multline}
  \Delta^\text{ph, (b)}_{\text{xc}\;i} = \frac{1}{2}\frac{Z^1_i}{\sum_{j}Z^0_j}
  \sum_{j l} \sum_{\lambda,\bq}
  \left|g^{jl}_{\lambda,\bq}\right|^2
  \frac{\Delta_{j}\Delta_{l}^*+
  \Delta_{j}^*\Delta_{l}}{E^2_{j} E_{l}}
  \Bigg\{\frac{\xi_{j}}{E_{j}}
  \left[1-\frac{Y^1_j}{Y^0_j}\xi_{j}\right]
  \left[I(E_{j}, E_{l}, \Omega_{\lambda,\bq}) - 
  I(E_{j}, -E_{l}, \Omega_{\lambda,\bq})\right]
  \\
  - \left[\xi_{j}+\frac{Y^1_j}{Y^0_j}|\Delta_{j}|^2\right]
  \left[I'(E_{j}, E_{l}, \Omega_{\lambda,\bq}) - 
  I'(E_{j}, -E_{l}, \Omega_{\lambda,\bq})\right]
  \Bigg\}
  \\ - \frac{1}{2}
  \sum_{j}\sum_{\lambda,\bq}\frac{\Delta_i}{Y^0_iE_{j}} 
  \left|g^{ij}_{\lambda,\bq}\right|^2
  \Bigg\{
  \left[\frac{\frac{\beta}{2}\left(\Delta_{j}\Delta_{i}^*-\Delta_{j}^*\Delta_{i}\right)}
  {\tanh\left(\frac{\beta}{2} E_{i}\right)
  \cosh^2\left(\frac{\beta}{2} E_{j}\right)} + 
  2\frac{\Delta_{j}}{\Delta_i}\frac{\xi_i^2}{E_i}\right]
  \left[I(E_{i}, E_{j}, \Omega_{\lambda,\bq}) - 
  I(E_{i}, -E_{j}, \Omega_{\lambda,\bq})\right]
  \\ 
  + \left(\Delta_{j}\Delta_{i}^*+\Delta_{j}^*\Delta_{i}\right)
  \left[I'(E_{i}, E_{j}, \Omega_{\lambda,\bq}) - 
  I'(E_{i}, -E_{j}, \Omega_{\lambda,\bq})\right]
  \Bigg\}\,.
\end{multline}
\end{widetext}
It can be seen here that the non-linear gap equation~(\ref{eq:nlgap}), in general,
does not have the simple structure of a BCS-like gap equation and thus goes beyond
the simple picture of an effective interaction mediating the pairing.
However, this approach encompasses several problems. First, the resulting functionals
have extremely complicated analytical structures and are very hard to interpret in
simple physical terms. Furthermore, these functionals contain several divergences
and pathological behaviors that have to be taken care of. For the time being, we
restrict ourselves to using the {\it partially linearized} exchange-correlation
potential, leading to the BCS-type gap equation
\be
  \Delta_i = -\frac{1}{2}  \sum_j {\cal F}_{\text{Hxc}\;i,j} 
  \frac{\tanh\left(\frac{\beta}{2}E_j\right)}{E_j} \Delta_j
  \,,
\ee
where ${\cal F}_{\text{Hxc}\;i,j}$ are the linearized functionals defined in Eq.~(\ref{eq:deffxc}) and
derived in detail in the previous section. It turns out that superconducting gap functions obtained
with these functionals are in rather good agreement with experimental results (see II).

\section{Conclusions}
\label{sec:concl}

In this work we have developed a truly ab-initio approach to
superconductivity. No adjustable parameters appear in the theory. The
key feature is that the electron-phonon interaction and the Coulombic
electron-electron repulsion are treated on the same footing. This is
achieved within a density-functional-type framework. Three
``densities'': the ordinary electronic density, the superconducting
order parameter, and the diagonal of the nuclear $N$-body density
matrix were identified as suitable quantities to formulate the
density-functional framework. The formalism leads to a set of
Kohn-Sham equations for the electrons and the nuclei. The electronic
Kohn-Sham equations have the structure of Bogoliubov-de Gennes
equations but, in contrast to the latter, they incorporate normal and
anomalous xc potentials. Likewise, the Kohn-Sham equation describing
the nuclear motion contains, besides the bare nuclear Coulomb
repulsion, an exchange-correlation interaction.  The latter is an
$N$-body interaction, i.e., the nuclear Kohn-Sham equation is an
$N$-body Schr\"odinger equation. The exchange-correlation potentials
are functional derivatives of a universal functional
$F_\text{xc}[n,\chi,\Gamma]$ which represents the exchange-correlation
part of the free energy. Approximations for this functional were then
derived by many-body perturbation theory. To this end, the effective
nuclear interaction was expanded to second order in the displacements
from the nuclear equilibrium positions. By introducing the usual
collective (phonon) coordinates, the nuclear Kohn-Sham equation is
then transformed into a set of harmonic oscillator equations
describing independent phonons. These non-interacting phonons,
together with non-interacting but superconducting (Kohn-Sham)
electrons serve as unperturbed system for a G\"orling-Levy-type
perturbative expansion of $F_\text{xc}$. The electron-phonon
interaction and the bare electronic Coulomb repulsion, as well as some
residual exchange-correlation potentials are treated as the
perturbation. In this way, both Coulombic and electron-phonon
couplings are fully incorporated. 
The solution of the KS-Bogoliubov-de Gennes equation (or the KS gap equation together
with the normal-state Schr{\"o}dinger equation) fully determines the Kohn-Sham system.
Therefore, within the usual approximation to calculate observables from the
Kohn-Sham system, one can apply the full variety of expressions for physical
quantities, known from phenomenological Bogoliubov-de Gennes theory, also in 
the present framework. This approach was already successfully applied within
a semi-phenomenological parametrization of the xc-functional , e.g., 
to the specific heat\cite{gyorffy:98} and to the penetration depth \cite{szotek:00} of the 
cuprates.
It should further be emphasized that the formalism, developed in this paper,
is not restricted to perfect periodic systems. It was for this purpose that
we presented all formulae in terms of general quantum numbers. The formalism can,
in principle, be applied as well to inhomogeneous systems, containing, e.g., impurities or
surfaces, as to perfect periodic crystals.

In the succeeding paper (II) we will detail the numerical implementation
of this theory and present the first full-scale applications to simple metals.

\begin{acknowledgments}
S.M. would like to thank A. Baldereschi, G. Cappellini, and G. Satta for useful discussions.
This work was partially supported by the INFM through the Advanced Research Project (PRA) UMBRA, 
by a supercomputing grant at Cineca, (Bologna, Italy) through the Istituto Nazionale di 
Fisica della Materia (INFM), and by the 
Italian Ministery of Education, through a 2004 PRIN project. Financial
support by the Deutsche Forschungsgemeischaft within the program SPP 1145, by the EXC!TiNG 
Research and Training Network of the European Union, and by the NANOQUANTA Network of
Excellence is gratefully acknowledged.
\end{acknowledgments}

\begin{appendix}
\section{On the electron-phonon coupling}
\label{appendix-1}
In this appendix we discuss the electron-phonon coupling potential
which appears in the phononic exchange-correlation 
terms. If we decomposed the Hamiltonian 
into the ionic Kohn-Sham Hamiltonian and the electronic Kohn-Sham
Hamiltonian as the reference system and the rest as perturbation, 
this perturbation would include the bare electron-ion interaction.
Clearly, the use of the bare vertex, i.e.
the gradient of the bare nuclear potential with respect to the 
nuclear positions, would yield an unphysical electron-phonon interaction. This bare vertex will
be screened by the conduction electrons. This screening could be taken 
care of by a diagrammatical resummation \cite{pLeeuwen}. 

Here we will sketch a different approach which directly generates the screened 
coupling potential. A natural coupling potential in the context
of DFT is the gradient with respect to the nuclear positions of the effective Kohn-Sham potential
within the Born-Oppenheimer approximation.
This is also exactly the quantity which is obtained from the 
standard DFT electron-phonon calculations based on linear response
theory with respect to small lattice distortions.
\begin{multline}
  \nabla_{\underline{\bR}} v_{\rm s,\underline{\bR}}(\br) = 
  \nabla_{\underline{\bR}} v^{\rm latt,}_{\underline{\bR}}(\br) \\
  + \mdint{r'}{r''} f_{\rm Hxc}(\br,\br') {\cal X}(\br',\br'')
  \left[ \nabla_{\underline{\bR}} v^{\rm latt}_{\underline{\bR}}(\br'')
  \right] 
\end{multline}
${\cal X}(\br,\br')$ denotes the linear density-density response 
function and
\be
  f_{\rm Hxc}(\br,\br') = \frac{\delta}{\delta n(\br')} \left\{ v_{\rm H}[n](\br) + v_{\rm xc}[n](\br) \right\} \,.
\ee

We are going to outline an approach which generates
exactly this gradient of the effective Kohn-Sham potential
as the coupling potential. The effective Kohn-Sham Hamiltonian (\ref{KSNa})
for the nuclei gives rise to a set of equilibrium coordinates
and phonon eigenstates. It can then
(up to harmonic order) be written as Eq.~(\ref{eq:h_phonon}).
The equilibrium positions $\underline{\bR_0}$ can be employed to define an
electronic BO Hamiltonian with a lattice potential
referring to these coordinates 
\be
  \label{H_BO}
  \op{H}^\text{BO} = \op{T}^\text{e} + \op{U}^\text{ee} + \op{V}^\text{latt}_{\underline{\bR_0}} \,.
\ee
This BO Hamiltonian, without the electron-phonon coupling, gives rise 
to the electronic density $n_{\underline{\bR_0}}$.
We now add and subtract all Hartree and exchange-correlation terms
as well as the BO lattice potential to the full Hamiltonian.
\bea
\op{H} &=\phantom{+}& (\op{T}^\text{n} + \op{U}^\text{nn} + \op{V}^\text{n}_\text{Hxc}) \nonumber \\
       &+& ( \op{T}^\text{e} + \op{V}^\text{latt}_{\underline{\bR_0}} 
       + \op{V}^\text{e}_\text{Hxc} + \op{\Delta}^\text{e}_\text{Hxc} ) \nonumber \\
       &+& \op{U}^\text{ee} + ( \op{U}^\text{en} - \op{V}^{\text{latt}}_{\underline{\bR_0}}) \nonumber \\
       &-& \op{V}^\text{n}_\text{Hxc}
       - \op{V}^\text{e}_\text{Hxc} - \op{\Delta}^\text{e}_\text{Hxc} \,.
\eea
The first three terms of this operator represent the nuclear Kohn-Sham
Hamiltonian. Assuming that the equilibrium density $n(\br)$ entering the
functional $v^{\rm e}_{\rm Hxc}[n](\br)$ will be close to the equilibrium
density $n_{\underline{\bR_0}}(\br)$ resulting from the BO Hamiltonian
(\ref{H_BO}) we can expand the Hartree exchange-correlation
potential around the BO density,
\begin{multline}
  v^{\rm e}_{\rm Hxc}[n](\br) = \\
  v^{\rm e}_{\rm Hxc}[n_{\underline{\bR_0}}](\br) 
  + \mint{r'} f_{\rm Hxc}[n_{\underline{\bR_0}}](\br,\br') 
  \delta n(\br') \,.
\end{multline}
where the small density change $\delta n(\br)$ is induced by the difference of the full
electron-ion interaction and the BO potential. The density change can, 
in principle, be
calculated via linear response to that perturbation. 
We expect this density change to be close to
\be
  \delta n(r') = \mint{r''} {\cal X}(\br',\br'')
  \left[ \nabla_{\underline{\bR}} v^{\rm latt}_{\underline{\bR}}(\br'') \right]
  \,.
\ee
If we keep only the BO part of the electronic Hartree exchange-correlation potential,
i.e., the term stemming from $n_{\underline{\bR_0}}(\br)$, in the electronic 
Kohn-Sham Hamiltonian, we can combine the remainder with the 
electron-ion interaction, and can identify (up to first order)
\begin{multline}
  \sum_i \frac{Z}{|\br-\bR_i|} - \sum_i \frac{Z}{|\br-\bR_{0,i}|} \\
  + \mint{r'} f_{\rm Hxc}[n_{\underline{\bR_0}}](\br,\br') \delta n(\br')
  \approx \nabla_{\underline{\bR}} v_{\rm s,BO}(\br) .
\end{multline}
This is the desired result, which allowed us to use the electron-phonon
couplings, determined by linear response calculations, as the 
coupling potentials in our Kohn-Sham perturbation theory.

\end{appendix}



\end{document}